\documentstyle[times,graphics,epsfig,amsmath,amssymb,psfrag,natbib,color,rotate,rotating]{mn2e}

\newcommand{\be}{\begin{equation}}
\newcommand{\e}{\end{equation}}
\newcommand{\bear}{\begin{eqnarray}}
\newcommand{\ear}{\end{eqnarray}}

\def\xh1{x_{H~{\sc I}\,}}
\def\x^i{x^{i}_{{\rm H~{\sc{i}}}}\,}
\def\xb{\bar{x}_{{\rm H~{\sc{i}}}}}
\def\Ph1{P_{{H~{\sc{i}}}}}
\def\eh1{\eta_{{H~{\sc{i}}}}}

\def\a{{\mathcal A}}

\def\HI{H{\sc i}\,}
\def\HII{H{\sc ii}\,}
\def\P{\hat{P}}

\def\k{{\bf k}}

\def\Tb{{T_{\rm b}}}
\def\tTb{\tilde{T}_{\rm b}}
\def\k{{\bf \textit{k}}}
\def\x{{\bf \textit{x}}}
\def\a{{\bf \textit{a}}}
\def\snr{{\rm SNR}}
\def\cov{{\mathcal{C}}}
\def\dcov{{c}}

\begin{document}
\title[EoR 21-cm power spectrum error-covariance]{Statistics of the epoch of reionization 21-cm signal -- I. Power spectrum error-covariance}
\author[Mondal, Bharadwaj and Majumdar]{Rajesh Mondal$^{1, 2}$\thanks{rm@phy.iitkgp.ernet.in},
Somnath Bharadwaj$^{1, 2}$\thanks{somnath@phy.iitkgp.ernet.in}, Suman Majumdar$^3$
\\
$^1$ Department of Physics, Indian Institute of Technology Kharagpur, Kharagpur 721302, India\\
$^2$ Centre for Theoretical Studies, Indian Institute of Technology Kharagpur, Kharagpur 721302, India\\
$^3$ Department of Astronomy and Oskar Klein Centre, AlbaNova, 
Stockholm University, SE-106 91 Stockholm, Sweden}
\date{Accepted 2015 November 23.  Received 2015 November 16; in original form 2015 August 4}

\pagerange{\pageref{firstpage}--\pageref{lastpage}} \pubyear{2015}

\maketitle

\label{firstpage}

\begin{abstract}
  The non-Gaussian nature of the epoch of reionization (EoR) 21-cm
  signal has a significant impact on the error variance of its power
  spectrum $P(\k)$. We
  have used a large ensemble of semi-numerical simulations and an
  analytical model to estimate the effect of this non-Gaussianity on
  the entire error-covariance matrix $\cov_{ij}$. Our analytical model
  shows that $\cov_{ij}$ has contributions from two sources. One is
  the usual variance for a Gaussian random field which scales
  inversely of the number of modes that goes into the estimation of
  $P(\k)$. The other is the trispectrum of the signal.  Using the
  simulated 21-cm signal ensemble, an ensemble of the randomized
  signal and ensembles of Gaussian random ensembles we have quantified
  the effect of the trispectrum on the error variance $\cov_{ii}$. We
  find that its relative contribution is comparable to or larger than
  that of the Gaussian term for the $k$ range $0.3 \leq k \leq 1.0
  \,{\rm Mpc}^{-1}$, and can be even $\sim 200$ times larger at $k
  \sim 5\, {\rm Mpc}^{-1}$.  We also establish that the off-diagonal
  terms of $\cov_{ij}$ have statistically significant non-zero values
  which arise purely from the trispectrum. This further signifies that
  the error in different $k$ modes are not independent.  We find a
  strong correlation between the errors at large $k$ values ($\ge 0.5
  \,{\rm Mpc}^{-1}$), and a weak correlation between the smallest and
  largest $k$ values. There is also a small anti-correlation between
  the errors in the smallest and intermediate $k$ values. These
  results are relevant for the $k$ range that will be probed by the
  current and upcoming EoR 21-cm experiments.
\end{abstract}
\section{Introduction}
\label{sec:intro}
The epoch of reionization (EoR) is one of the least known but
important periods in the history of our Universe. During this epoch
the diffused hydrogen in our universe gradually changed from neutral
to ionized. Our current knowledge about this epoch is very
limited. The measurements of the Thomson scattering optical depth of
the cosmic microwave background (CMB) photons from the free electrons
in the intergalactic media (IGM) (e.g. see
\citealt{komatsu11,planck14,planck15} etc.) and the observations of the
Lyman-$\alpha$ absorption spectra of the high redshift quasars
(e.g. see\citealt{becker01,fan03,white03,goto11,becker15} etc.) suggest
that this epoch was probably extended over a wide redshift range $6
\lesssim z \lesssim 12$ (see
e.g. \citealt{mitra11,mitra15,mitra13,robertson13,robertson15}). However,
many fundamental issues such as the characteristics of the major
ionizing sources, the precise duration and timing of reionization and
the topology of the neutral hydrogen (\HI) distribution etc. cannot be
resolved using these indirect observations.

Observation of the redshifted \HI 21-cm signal which provides a direct
window to the state of the hydrogen in the IGM is a very promising
probe of the EoR. There is a considerable effort underway to detect
the EoR 21-cm signal through radio interferometry using e.g. the
{GMRT\footnote{http://www.gmrt.ncra.tifr.res.in}} \citep{paciga13},
{LOFAR\footnote{http://www.lofar.org/}} \citep{haarlem13,yatawatta13},
{MWA\footnote{http://www.haystack.mit.edu/ast/arrays/mwa/}
  \citep{bowman13,tingay13,dillon14}} and
{PAPER\footnote{http://eor.berkeley.edu/}}
\citep{parsons14,ali15,jacobs14}. Apart from these first-generation
radio interferometers, the detection of this signal is also one of the
key science goals of the future telescopes such as the
{SKA\footnote{http://www.skatelescope.org/}
  \citep{mellema13,koopmans15}} and
{HERA\footnote{http://reionization.org/} \citep{furlanetto09}}. The
\HI 21-cm signal is expected to be very weak ($\sim 4-5$ orders in
magnitude) compared to the enormous amount of foreground emission,
from our own galaxy and the extragalactic sources, in which it is
buried
\citep{dimatteo02,gleser08,ali08,jelic08,bernardi09,ghosh12,pober13,moore13,moore15}.
Mainly these foregrounds, the system noise \citep{morales05,mcquinn06}
and the other sources of calibration errors together have kept the
cosmologists at bay from detecting this signal and till today only a
rather weak upper limit on it have been obtained
\citep{paciga13,dillon14,parsons14,ali15}. Due to these obstacles, it
is anticipated that the first detection of the signal will be through
statistical estimators such as the variance \citep{patil14} and the
power spectrum \citep{pober14}, where one adds up the signal optimally
to enhance the signal-to-noise ratio (SNR).
      
Any statistical estimation of a signal comes with an intrinsic
uncertainty of its own, which arises because of the uncertainties in
the signal across its different statistically independent
realizations. In cosmology, this uncertainty is more commonly known as
the \emph{cosmic variance} (in other words this is the uncertainty due
to the fact that we have only one universe to estimate the
signal). Apart from the cosmic variance there will be uncertainties
due to the sensitivity of the instrument as well (e.g. system noise,
non-uniform baseline distribution etc.). It is necessary to quantify
the different possible uncertainties in these measurements to
correctly interpret the signal once it has been detected. If the EoR
21-cm signal had the nature and properties similar to a Gaussian
random field, the estimation of its cosmic variance would have been
very straight forward, as it scales as the square root of the number
of independent measurements. Almost all studies
(e.g. \citealt{morales05, mcquinn06, beardsley13, jensen13, pober14,
  koopmans15} etc.) that have been undertaken to quantify the
detectability of the EoR 21-cm power spectrum using different
instruments such as the MWA, LOFAR, PAPER, SKA etc. assume the signal
to be a Gaussian random field while estimating its cosmic
variance. This can be a reasonably good assumption at large length-scales 
during the early phases of reionization when the \HI is
expected to trace the underlying dark matter distribution. However,
during the intermediate and the later stages of the reionization, the
signal only appears from the neutral hydrogen located on the periphery
of the ionized (\HII) regions which are gradually growing both in
number and size. This makes the redshifted 21-cm signal from the later
stages of EoR highly non-Gaussian.

The statistics of a Gaussian random field is completely specified by
its power spectrum, whereas the higher order statistics like the
bispectrum \citep{bharadwaj05} and the trispectrum are also important
for a highly non-Gaussian field like the EoR 21-cm signal. Though the
power spectrum itself cannot capture the non-Gaussian nature of the
signal, the non-Gaussianity however will significantly affect its
error estimates (i.e. cosmic variance). This has been demonstrated in
a recent work by \citet{mondal15} using a large ensemble of simulated
EoR 21-cm signal. \citet{mondal15} have shown that for a fixed
observation volume, it is not possible to obtain an SNR above a certain
limiting value, even when one increases the number of Fourier modes
that goes into the estimation of the power spectrum. The analytical
model for the cosmic variance proposed in this work further indicates
that this limiting value of the SNR is directly related to the
trispectrum of the signal and the total survey volume under
consideration.

In this follow-up work on \citet{mondal15}, we extend their analytical
model to derive a generic expression for the entire error covariance
matrix of the binned 21-cm power spectrum. Using a large number of
realizations of the simulated 21-cm signal from EoR we further attempt
to quantify the error covariance of its power spectrum. We also
interpret it in the light of this improved analytical model. Since,
this study is limited by the finite number of realizations of the
signal, thus we further check the statistical significance of this
error covariance. Besides this, the entire analysis of this paper is
based on the numerical simulations of 21-cm signal which have a finite
comoving volume. We therefore test the convergence of our results by
increasing our simulation volume. Finally, we have tried to extract
the trispectrum of the signal from the non-Gaussian component of the
error covariance of the power spectrum. It is also important to note
that the nature of the results and the analytical model that we have
presented here is not limited only to the EoR 21-cm signal but can be
applied to the analysis of any non-Gaussian cosmological signal such
as the galaxy redshift surveys
\citep{feldman94,neyrinck11,mohammed14,caron14}.

The structure of this paper is as follows. Starting from the basic
definition of the 21-cm brightness temperature fluctuations we derive
the expressions for the power spectrum and the trispectrum of the EoR
redshifted 21-cm signal in Section \ref{sec:model}. We next derive the
error covariance of the binned power spectrum estimator and also show
its relation to the trispectrum in Section \ref{sec:covar}. Section
\ref{sec:sim} describes the semi-numerical simulations that we have
used to generate the realizations of the EoR 21-cm signal. In Section
\ref{sec:ensmbl}, we describe about the reference ensembles which are
used to interpret the results. In Section \ref{sec:results}, we describe 
our results i.e. the estimated error covariance of the power 
spectrum from the simulated data. Finally, in Section \ref{sec:summary}, 
we discuss and summarize our results.

Throughout this paper, we have used the \textit{Planck}+WP best-fitting 
values of cosmological parameters $\Omega_{\rm m0}=0.3183$,
$\Omega_{\rm \Lambda0}=0.6817$, $\Omega_{\rm b0}h^2=0.022032$, $h=0.6704$,
$\sigma_8=0.8347$ and $n_{\rm s}=0.9619$ \citep{planck14}.

\section{The power spectrum and the trispectrum}
\label{sec:model}
The EoR 21-cm signal is quantified through the brightness temperature
fluctuation
\begin{equation}
\delta \Tb (\x)=\Tb (\x)- \bar{T}_{\rm b} \,.
\end{equation}
In this paper we are interested in the statistical properties of
$\delta \Tb (\x)$ which is assumed to be a statistically homogeneous 
random field. The two point statistics of $\delta \Tb
(\x)$ is quantified through the two-point correlation function $\xi$
which is defined as
\begin{equation}
\langle \delta \Tb (\x_1) \, \delta \Tb (\x_2) \rangle = \xi (\x_1,\x_2) 
\label{eq:2point}
\end{equation}
where the angular brackets $\langle . . .\rangle$ denote an ensemble
average over many statistically independent realizations of $\delta
\Tb (\x)$. It follows from statistical homogeneity that the two-point
correlation function  is  invariant if we apply a displacement $\a$ to
both $\x_1$ and $\x_2$, or equivalently $\xi$   
 depends only on $\x_{21}=\x_2-\x_1$ the relative
displacement vector between the two points $\x_1$ and $\x_2$
\begin{equation} 
\xi(\x_1,\x_2) =  \xi (\x_1+\a,\x_2+ \a)  =\xi(\x_{21}) \,.
\label{eq:hom1}
\end{equation}

The EoR 21-cm signal is not statistically isotropic due to redshift space
distortion \citep{bharadwaj04}. While several works have attempted to quantify
this anisotropy \citep{majumdar13,jensen13,
    shapiro13,majumdar14,ghara14, fialkov15,majumdar15},
in this work we  only consider $\xi(x_{21})$, which is the
monopole (isotropic) component of $\xi(\x_{21})$.

We now consider the four point statistics (see e.g.
  equation 35.3 of \citealt{peebles1980})
\begin{align}
&\langle \delta \Tb (\x_1) \, \delta \Tb (\x_2) \delta \Tb (\x_3) \, \delta
\Tb (\x_4) \rangle =
\xi(x_{12}) \xi(x_{34}) \nonumber \\
& + \xi(x_{13}) \xi(x_{24})+ \xi(x_{14}) \xi(x_{23}) + \eta (\x_1, \x_2, \x_3, \x_4)
\label{eq:4point}
\end{align}
where the (reduced) four-point correlation function $\eta$ quantifies
the excess over the product of $\xi$s. Here, statistical homogeneity
implies that $\eta$  is  invariant if we apply a displacement $\a$ to
$\x_1$, $\x_2$, $\x_3$ and $\x_4$ i.e.
\begin{equation}
\eta (\x_1, \x_2, \x_3, \x_4) = \eta (\x_1+\a, \x_2+\a, \x_3+\a, \x_4+\a) 
\label{eq:hom2}
\end{equation}
or equivalently $\eta$  depends only on three relative  displacement vectors  
\begin{equation}
\eta (\x_1, \x_2, \x_3, \x_4) = \eta (\x_{21}, \x_{31}, \x_{41}) \,.
\label{eq:hom3}
\end{equation}

It is convenient to use the Fourier representation considering a cubic
comoving volume $V$ with periodic boundary conditions. We then have
\begin{equation}
\Tb (\x) = \frac{1}{V} \sum^{}_{\k} {\rm e}^{i\k 
\cdot \x} \, \tTb ({\k})
\label{eq:ft}
\end{equation}
where $\tTb ({\k})$ is the Fourier transform of 
$\Tb ({\x})$. Note that the wave vector  $\k$ assumes  both positive and 
negative values, however these are not independent as we have the relation
$\tTb ^{*}({\k})=\tTb (-{\k})$ which holds for the Fourier transform of a real
quantity. Further, we can equally well interpret $\tTb ({\k})$ 
as the Fourier transform of $\delta \Tb
({\x})$ for all values of $\k$ barring $k=0$.

We first consider the two-point statistics. Incorporating the Fourier
representation equation (\ref{eq:ft}) in equation (\ref{eq:2point}), we have 
\begin{equation}
\xi (\x_1,\x_2) = \frac{1}{V^2} \sum^{}_{\k_1 ,\k_2} e^{i(\k_1 
\cdot \x_1 + \k_2 \cdot \x_2) } \, \langle  \tTb ({\k_1}) \tTb ({\k_2})
\rangle 
\label{eq:ft1}
\end{equation}
We see that the r.h.s. picks up  an extra phase factor $Q=e^{i(\k_1  + \k_2)
  \cdot {\bf a}} $ 
if we apply a  displacement  ${\bf a}$ to   both $\x_1$ and $\x_2$. However, the
assumption of  statistical homogeneity (equation \ref{eq:hom1})  requires
equation (\ref{eq:ft1}) to be invariant  under such a displacement.  This implies
that $ \langle \tTb ({\k_1}) \tTb ({\k_2}) \rangle$  has non-zero values 
only when $\k_1  + \k_2=0$  for which  $Q=1$, and it is zero when  $\k_1  +
\k_2 \neq 0$, We than have 
\begin{equation}
\langle \tTb(\k_1) \, \tTb(\k_2) \rangle = \delta_{\k_1+\k_2,0} V P(\k_1)
\label{eq:pk1}
\end{equation}
where the Konecker delta $\delta_{\k_1+\k_2,0}$ 
is 1 if $\k_1+\k_2 = 0$ and 0 otherwise. Here 
the power spectrum  $P(\k)=P(k)$ is defined as 
\begin{equation}
P(\k)=V^{-1} \langle \tTb(\k) \, \tTb(-\k) \rangle \,.
\label{eq:pkdef}
\end{equation}
Using equation (\ref{eq:pk1}) in  equation (\ref{eq:ft1}), we have 
\begin{equation}
\xi (\x_1,\x_2) = \frac{1}{V^2} \sum^{}_{\k_1 ,\k_2} e^{i(\k_1 
\cdot \x_1 + \k_2 \cdot \x_2) } \, \times V \delta_{\k_1+\k_2,0}  P(\k_1)
\label{eq:ft2}
\end{equation}
whereby we see that 
the power spectrum is the Fourier transform of
the two-point correlation function
\begin{equation}
\xi(\x_{21}) = \frac{1}{V} \sum^{}_{\k} e^{i\k 
\cdot \x_{21}} \, P(\k) \,.
\label{eq:pk2}
\end{equation}

Proceeding in exactly the same manner for the four-point statistics
(equation \ref{eq:4point}), statistical homogeneity (equation \ref{eq:hom2}) 
requires that 
\begin{align}
&\langle \tTb (a) \tTb (b) \tTb (c) \tTb (d) \rangle = V^2 [ \, \delta_{a+b,0} \,  \delta_{c+d,0} \, P(a) P(c)
\nonumber \\
&+ \delta_{a+c,0} \delta_{b+d,0} P(a) P(b)
+ \delta_{a+d,0} \delta_{b+c,0} P(a) P(b)] 
\nonumber \\
&+ V \delta_{a + b + c + d,0}  \, T (a,b,c,d)
\label{eq:tk1}
\end{align}
where we have used the notation $\tTb (a) \equiv \tTb ({\k_a})$. Here 
the trispectrum
$T (\k_a,\k_b,\k_c,\k_d)$ is the Fourier transform of the four-point
correlation function
\begin{align}
\eta(\x_1,\x_{2},\x_{3},\x_{4}) &= \frac{1}{V^4} \sum_{\k_1,\k_2,\k_3,\k_4}
e^{i(\k_1 \cdot \x_1 +\k_2 \cdot \x_{2} + \k_3 \cdot \x_{3} + \k_4 \cdot \x_{4})} \, 
\nonumber \\
&\times V \delta_{\k_1+\k_2+\k_3+\k_4,0} T (\k_1,\k_2,\k_3,\k_4)\,.
\label{eq:tk2}
\end{align}
Note that equation (\ref{eq:tk2}) for the four point statistics is exactly
analogous to equation (\ref{eq:ft2}) which has been discussed earlier for the
two-point statistics. We can also carry out the sum over $\k_1$ and express 
equation (\ref{eq:tk2}) as
\begin{align}
\eta(\x_{21},\x_{31},\x_{41}) &= \frac{1}{V^3} \sum_{\k_2,\k_3,\k_4}
e^{i(\k_2 \cdot \x_{21} + \k_3 \cdot \x_{31} + \k_4 \cdot \x_{41})} \, 
\nonumber \\
&\times T (-\k_2-\k_3-\k_4,\k_2,\k_3,\k_4)\,.
\label{eq:tk3}
\end{align}

The entire analysis of this paper is based on numerical simulations
which have a finite comoving volume $V$. The various factors of $V$
that appear in equations (\ref{eq:pk1}), and (\ref{eq:pk2}) -- 
(\ref{eq:tk2}) leave one wondering whether the power spectrum and
particularly the trispectrum would vary if the volume $V$ were
changed. To address this, we consider the limit $V \rightarrow
\infty$. In this limit the power spectrum
\begin{equation}
[P(\k)]_{\infty}=\int \xi(\x_{21})  e^{-i \k \cdot \x_{21}}\, d^3 x_{21}
\end{equation}
and  the trispectrum 
\begin{align} 
&[T (-\k_2-\k_3-\k_4,\k_2,\k_3,\k_4)]_{\infty} =
   \int \eta(\x_{21},\x_{31},\x_{41}) \times \nonumber \\
  &e^{-i (\k_2 \cdot \x_{21}+\k_3 \cdot \x_{31}+\k_4 \cdot \x_{41})} \,
  d^3 x_{21}\, d^3 x_{31}\, d^3 x_{41}
\end{align}
have finite, well defined values provided the integrals
\begin{equation}
\int \mid \xi(\x_{21}) \mid \, d^3 x_{21}
\label{eq:conv1}
\end{equation}
and 
\begin{equation}
\int \mid \eta(\x_{21},\x_{31},\x_{41}) \mid \, d^3 x_{21} \, d^3 x_{31} \,
d^3 x_{41}
\label{eq:conv2}
\end{equation}
respectively converge. 

We have assumed that $\xi(\x_{21})$ and
$\eta(\x_{21},\x_{31},\x_{41})$ fall sufficiently rapidly at large
separations so that the integrals in equations (\ref{eq:conv1}) and
(\ref{eq:conv2}) both converge.  The limiting power spectrum
$[P]_{\infty}$ and trispectrum $[T]_{\infty}$ then have finite,
well defined values, and the simulated $P$ and $T$ would
respectively converge to $[P]_{\infty}$ and $[T]_{\infty}$ if the
simulation volume $V$ were increased. In our analysis we assume that
our simulations cover a sufficiently large volume of the universe
whereby the simulated power spectrum and trispectrum are respectively
sufficiently close to $[P]_{\infty}$ and $[T]_{\infty}$ for the
$k$ range of our interest, and the simulated values would not change
significantly if the volume $V$ were increased further.

\section{The error-covariance of the power spectrum}
\label{sec:covar}
The question here is `How accurately can we estimate the power
spectrum from a given EoR data?'. In general, any observation will
yield a combination of the EoR signal and instrumental noise, assuming
that the foregrounds have been completely subtracted out.  In this
analysis, we only consider the statistical errors which are inherent to
the EoR signal, and we do not consider the instrumental noise.  The
statistical errors which we have considered here are usually referred
to as the \emph{cosmic variance}.

We consider the binned power spectrum estimator $\hat{P}(k_i)$ which, 
for the $i$~th bin, is defined as 
\begin{equation}
\hat{P}(k_i)= \frac{1}{N_{k_i} V} \sum_{\k}  \tTb (\k) \,\tTb (-\k) \, ,
\label{eq:est}
\end{equation}
where $\sum_{\k}$, $N_{k_i}$ and $k_i$ respectively refer to the sum,
the number and the average comoving wavenumber of all the Fourier
modes in the $i$~th bin.  The bins here are spherical shells of width
$\Delta k_i$ in Fourier space.  We have used logarithmic binning which 
essentially implies that $\Delta k_i \, (\propto k_i) $ will vary from
bin to bin.  As the modes $\k$ and $-\k$ do not provide independent
estimates of the power spectrum, we have restricted the sum to the
upper half of the spherical shell 
which has volume $(2 \pi) \, k_i^2 \, \Delta
k_i$ in $\k$ space. To calculate $N_k$,  the number of Fourier modes in this
volume, we note that  the different wave vectors $\k$ are all equally
spaced at a separation of  $(2 \pi)/V^{1/3}$ in $\k$ space. We consequently have   
\begin{equation}
N_{k_i} \approx \, \frac{ (2 \pi) {k_i}^2 \Delta {k_i}}{\left[(2\pi)/V^{1/3}\right]^3}
= \frac{V}{(2 \pi)^2} \times k_i^2  \Delta {k_i}
\label{eq:n_k}
\end{equation}
which we use to estimate $N_{k_i}$.

The ensemble average  of the estimator gives the bin-averaged 
power spectrum 
\begin{equation}
\langle \hat{P}({k_i}) \rangle  = \bar{P} ({k_i}) =
\frac{1}{N_{{k_i}}} \sum_a P(a) \, .
\label{eq:pk}
\end{equation}

The  error-covariance  of the power spectrum estimator 
\begin{equation}
\cov_{ij} = \langle [\hat{P}({k_i})- \bar{P} ({k_i})] \,
[\hat{P}({k_j})- \bar{P} ({k_j})] \rangle 
\label{eq:cov1a}
\end{equation}
is the quantity of interest here. This  can also be written as 
\begin{equation}
\cov_{ij} = [\langle \hat{P}({k_i}) \, \hat{P}({k_j}) \rangle] - 
\bar{P} ({k_i})\, \bar{P} ({k_j}) \,.
\label{eq:cov2a}
\end{equation}
and the term in the square brackets $[...]$ of equation (\ref{eq:cov2a}) can be
expressed as 
\begin{equation}
[...]=
\frac{1}{N_{k_i} N_{k_j} V^2} \sum_{a \in i,b \in j} \langle \tTb (a) \tTb(-a)
\tTb(b) \tTb (-b) \rangle \,.
\label{eq:cov3a}
\end{equation}
Using eq. \eqref{eq:tk1} to simplify eq. \eqref{eq:cov3a}  we can express the
error covariance as
\begin{equation}
\cov_{ij} = \frac{\overline{P^2}(k_i)}{N_{k_i}}
 \, \delta _{ij} \, + \frac{\bar{T} (k_{i},k_{j})}{V}\,,
\label{eq:cov1}
\end{equation}
where
\begin{equation}
\overline{P^2}(k_i)=\frac{1}{N_{k_i}} \sum_a P^2(a) 
\end{equation}
is the   square of the power spectrum  averaged over the $i$~th bin,
and 
\begin{equation}
\bar{T} (k_i,k_j) = \frac{1}{N_{k_i} N_{k_j}} \sum_{a \in i,b \in j}  T (a,-a,b,-b)
\label{eq:tri}
\end{equation}
is the average trispectrum where  $\k_a$ and $\k_b$ are 
summed  over the  $i$~th and the $j$~th bins  respectively. 

We first discuss the results expected for a  Gaussian random field for 
which the  trispectrum is zero. In this case we can use 
equation (\ref{eq:n_k}) to express the covariance matrix as  
\begin{equation}
\cov_{ij} = \frac{1}{V} \left[ \frac{(2 {\pi})^2
 \, \overline{P^2}(k_i)}{k_i^2 \, \Delta k_i} \right] \,
 \, \delta _{ij} \,.
\label{eq:covg}
\end{equation}

The first  point here is that the covariance matrix is diagonal.
This implies that the errors in the   different bins  are  uncorrelated. 
The second point is that the covariance matrix scales as 
$\cov_{i j} \propto (V \, \Delta k_i)^{-1}$
 if we increase the observational volume $V$ or the bin width 
$\Delta k_i$.

It is possible to interpret the 
diagonal elements   $\cov_{ii}$ as the error variance 
$\cov_{ii}=[\delta P(k_i)]^2$ for  the power spectrum. 
We can then express the error in the estimated power spectrum as 
\begin{equation}
\delta P (k_i) = \sqrt{\frac{(2 \pi)^2 \,  \overline{P^2}(k_i)}
{V k_i^2 \, \Delta k_i}} \, 
\label{eq:var}
\end{equation}
which  is analogous to the error estimate in the context of 
galaxy redshift surveys (e.g. equation 11.119 of \citealt{dodelson_b03}). 
We see that the error falls as $\delta P (k_i) \propto 1/\sqrt{V}$ 
if we increase the observational volume. 
 For a fixed observational volume, we expect 
the error to fall  as $\delta P (k_i) \propto 1/\sqrt{ \Delta k_i}$ 
until it reaches  a minimum value which is achieved when 
all the Fourier modes are combined into a single bin. 

The EoR  signal becomes increasingly non-Gaussian as the reionization 
proceeds. This manifests itself as a non-zero trispectrum in 
the error-covariance (equation \ref{eq:cov1}) which can be expressed as 
\begin{equation}
\cov_{ij} = \frac{1}{V} \left[ \left( \frac{(2 {\pi})^2
 \, \overline{P^2}(k_i)}{k_i^2 \, \Delta k_i} \right) \,
 \, \delta _{ij} \, 
+  \bar{T} (k_{i},k_{j})\ \right] \,.
\label{eq:cov}
\end{equation}

The covariance matrix still retains the $1/V$ 
dependence,   similar to the Gaussian random field discussed 
earlier.  Consequently we still expect 
the errors in the estimated power spectrum to  go down
as  $1/\sqrt{V}$ if the observational volume is increased. 
However, the covariance matrix now has two major differences
from that of a Gaussian random field. 

The first difference is that the covariance matrix is no longer 
diagonal.   The average trispectrum  $\bar{T} (k_{i},k_{j})$
quantifies the correlation between the EoR signal in two 
different bins ($i$ and $j$).  
The off-diagonal elements of the covariance matrix 
($\cov_{i j}=\bar{T} (k_{i},k_{j})/V$) quantifies the 
correlation between the errors in the power spectrum estimated 
in the $i$ and $j$ bins respectively. 

The second difference is that the diagonal terms of the 
covariance matrix  deviate from the 
$\cov_{ii} \propto 1/ \Delta k_i$ 
behaviour predicted for a Gaussian random filed. 
For small bin-widths ($ \Delta k_i \, k^2_i  \ll
(2 {\pi})^2 \, \overline{P^2}(k_i)\,/ \bar{T} (k_{i},k_{i}) $), 
we  expect the error variance 
to fall as $\cov_{ii}\propto 1/ \Delta k_i$ as the bin-width 
$\Delta k_i$ is increased. The error variance  $\cov_{ii}$ 
saturates as the bin-width approaches $ \Delta k_i \, k^2_i  \approx
(2 {\pi})^2 \, \overline{P^2}(k_i)\,/ \bar{T} 
(k_{i},k_{i}) $, and it does not fall below  the 
limiting value 
$[\cov_{ii}]_l=\bar{T} (k_{i},k_{i})/V$ even if all the 
Fourier modes are combined into a single bin.

For a Gaussian random field, we expect the signal to noise ratio
$\snr_i=\bar{P}(k_i)/\delta P (k_i)$ to increase as $\snr_i
\propto \sqrt{N_{K_i}}$ if we increase the number of modes $N_{k_i}$
in the bin . The $\snr$, however, will saturate at a limiting value
$[\snr_i]_l=\bar{P}(k_i)/\sqrt{[\cov_{ii}]_l}$ when the EoR 21-cm
signal becomes non-Gaussian.  Semi-numerical simulations show
\citep{mondal15} that the $\snr_i \propto\sqrt{N_{k_i}}$ behaviour
only holds for small $\snr_i$, and $\snr_i$ saturates at a limiting
value $[\snr_i]_l$ as $N_{k_i}$ is increased. The limiting value
$[\snr_i]_l$ is found to decrease (i.e. $\cov_{ii}$ increases)
as reionization proceeds.

The expected $\cov_{ii} \propto 1/ \Delta k_i$ behaviour is a
consequence of the fact that the signal in the different Fourier modes
$\tTb (\k)$ is independent for a Gaussian random
field. The EoR signal at the different Fourier modes, however, become
correlated as ionized bubbles develop and the \HI signal becomes
non-Gaussian.  The trispectrum quantifies this correlation between the
signal at different Fourier modes.  The fact that $\cov_{ii}$
saturates and does not decrease beyond $[\cov_{ii}]_l$ even if we
increase $\Delta k_i$ is a consequence of the fact that we are not
adding independent information by increasing the number of Fourier
modes in the bin.

The trispectrum $T (\k_1,\k_2,\k_3,\k_4)$ is, in general 
(equation \ref{eq:tk1}), sensitive to correlations in both the amplitude 
and the phase of the signal at the different Fourier modes 
$\tTb (\k_1)$, $\tTb (\k_2)$, 
$\tTb (\k_3)$ and $\tTb (\k_4)$. 
The average trispectrum $\bar{T} (k_{i},k_{j})$ 
(equation \ref{eq:tri}).  which appears in our expression for the error 
covariance (equation \ref{eq:cov1}), however, depends only on the term 
$\langle \, \tTb (\k) \, \tTb^{*} (-\k) 
\, \tTb (\k^{'}) \, \tTb^{*} (-\k^{'}) 
\rangle $ which is insensitive to correlations in the 
phase of the modes $ \tTb (\k)$ and 
$\tTb (\k^{'})$. We therefore see that the 
error covariance $\cov_{ij}$ (equation \ref{eq:cov}) is only affected 
by the correlations in the amplitude of $\tTb (\k_i)$ 
and $\tTb (\k_j)$, it is insensitive to purely 
phase correlations between the signal at these two modes.

In summary of this section we note that the non-Gaussianity introduces
an extra term $\bar{T} (k_{i},k_{j})/V$ in the error
covariance (equation \ref{eq:cov}).  As a consequence the error variance
$\cov_{ii}$ for the binned power spectrum saturates at a limiting
value $[\cov_{ii}]_l$, it is not possible to decrease the error in the
estimated power spectrum beyond $\sqrt{[\cov_{ii}]_l}$ by increasing
the number of Fourier modes in the bin.  Further, the error covariance
matrix $\cov_{ij}$ is not diagonal.  The off-diagonal terms quantify
the correlations between the errors in the power spectrum estimated at
different bins.

\section{Simulating the EoR redshifted 21-cm signal}
\label{sec:sim}
\begin{figure*}
\centering
\includegraphics[width=1.01\textwidth, angle=0]{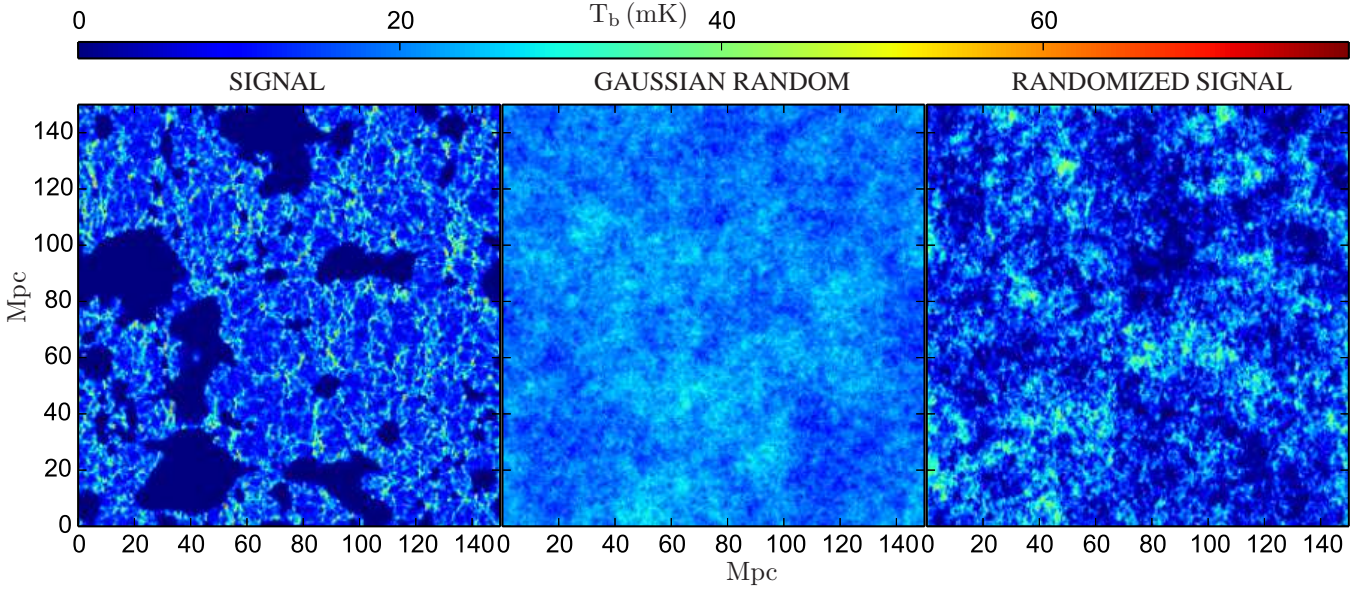}
\put(-241, -2.5){\large ${\rm Mpc}$}
\put(-513, 95){\large \begin{sideways}${\rm Mpc}$\end{sideways}}
\put(-293, 209){\large ${\rm T_b}\, ({\rm mK})$}
\put(-429, 183){\large{SIGNAL}}
\put(-291, 183){\large{GAUSSIAN RANDOM}}
\put(-133, 183){\large{RANDOMIZED SIGNAL}}

\caption{Two-dimensional sections through the simulated \HI brightness
  temperature maps for $\xb=0.5$ and $[150\,{\rm Mpc}]^3$ volume. The
  three panels each show a single realization drawn from the three
  different ensemble, Signal (left), Gaussian Random (middle) and
  Randomized Signal (right). The direction of redshift space
  distortion is with respect to a distant observer located along the
  horizontal axis.}
\label{fig:HI_map}
\end{figure*}
We have used \emph{semi-numerical} simulations to generate the EoR
redshifted 21-cm signal. These simulations consist of three main
steps. First, we use a particle mesh $N$-body code to generate the
dark matter distribution at the desired redshift.  We have run
simulations with two different comoving volumes $V_1=[150\, {\rm
  Mpc}]^3$ and $V_2 = [215\, {\rm Mpc}]^3$ using grids of size
$2144^3$ and $3072^3$, respectively. The spatial resolution $0.07 \,
{\rm Mpc}$ and the mass resolution $1.09 \times10^8 M_{\odot}$ is
maintained the same for both $V_1$ and $V_2$.  In the next step we
identify the mass and the location of collapsed haloes using the
standard Friends-of-Friends (FoF) algorithm \citep{davis85} with a
fixed linking length of $0.2$ times the mean inter-particle distance.
We have set the criterion that a halo should have at least $10$ dark
matter particles whereby we have a minimum halo mass of $1.09
\times10^9 M_{\odot}$

The final step generates the ionization map based on the excursion set
formalism of \citet{furlanetto04b}. The basic assumption here is that
the hydrogen traces the dark matter distribution and the dark matter
haloes host the sources which emit ionizing radiation.  It is assumed
that the number of ionizing photons emitted by a source is
proportional to the mass of the host halo, and it is possible to
achieve different values of the mass averaged \HI neutral fractions
$\xb$ by tuning this constant of proportionality.  Our simulation
closely follows \citet{choudhury09b} to generate the ionization map,
and the resulting \HI distribution is mapped onto redshift space to
generate 21-cm brightness temperature maps following
\citet{majumdar13}. The grid used to generate the ionization maps and
the 21-cm brightness temperature maps is eight times coarser than that
used for the $N$-body simulation.

The redshift evolution of $\xb$ is, at present, largely
unknown. Instead of assuming a particular model for $\xb(z)$, we have
fixed the redshift $z=8$ and run our simulations for different values
of $\xb$ at this fixed redshift. We have simulated \HI maps for $\xb$
values at an interval of $0.1$ in the range $1.0 \geq \xb \geq0.3$ in
addition to $\xb = 0.15$. For each simulation volume ($V_1$ and $V_2$)
and for each value of $\xb$, we have run $50$ independent simulations
to generate an ensemble of $50$ statistically independent realizations
of the 21-cm signal. We refer to this ensemble as the Signal Ensemble
(SE). The left-hand panel of Fig. \ref{fig:HI_map} shows a section
through one realization of the SE for $\xb=0.5$.  We have used the SE
to estimate the bin-averaged power spectrum $\bar{P} ({k_i})$ and the
error covariance matrix $\cov_{ij}$ for the two different simulation
volumes $V_1$ and $V_2$, and for the different $\xb$ values mentioned
earlier.

\section{Simulating reference ensembles}
\label{sec:ensmbl}
The previous section describes how we have estimated the power
spectrum error-covariance $\cov_{ij}$. In summary, we have constructed
an ensemble of $50$ statistically independent realizations of the
simulated EoR 21-cm signal and used this to estimate $\cov_{ij}$.  We
refer to this ensemble as the SE.  The question now
is `How do we interpret the estimated $\cov_{ij}$?'. We know that
for a Gaussian random field we expect: (A.) the diagonal terms to have
values as predicted by equation (\ref{eq:covg}), and (B.) the off-diagonal
terms to be zero.  We may interpret any deviation from this as arising
from non-Gaussianity, and then use these deviations to quantify the
contribution from the trispectrum in equation (\ref{eq:cov}).  While this
is straightforward in concept, several complications arise in
practice.

\subsection{The Randomized Signal Ensemble}
The first complication arises when we try to interpret the diagonal
terms $\cov_{ii}$. We expect these to have values as predicted by
equation (\ref{eq:covg}) if the signal were a Gaussian random field, and it
is possible to interpret deviations from this relation in terms of the
trispectrum which appears in equation (\ref{eq:cov}) when the signal
becomes non-Gaussian.  The problem arises because it is not possible
to use the SE to independently determine the value of
$\overline{P^2}(k_i)$ which appears in equation (\ref{eq:covg}).  We have
overcome this problem by constructing the Randomized Signal Ensemble
(RSE).

Each realization of RSE contains the signal drawn from all the $50$
realizations in SE. We have labeled all the modes in
  the simulation volume as $\k_1,\k_2,...$. Note that we are free to
  choose any arbitrary labeling scheme as long as it assigns an unique
  label to each distinct Fourier mode $\k$. The Fourier modes are
then divided into sets ${\mathcal A}_1=\{\k_1,\k_{51},\k_{101},...\}$,
${\mathcal A}_2=\{\k_2,\k_{52},\k_{102},...\}$,...  ${\mathcal
  A}_{50}=\{\k_{50},\k_{100},\k_{150},...\}$.  For the first
realization in RSE, the signal for all the modes in ${\mathcal A}_1$
is drawn from the first realization in SE (i.e. \, [SE]$_1$),
and the signal for all the modes in ${\mathcal A}_2$ is drawn from the
second realization in SE (i.e.  \, [SE]$_2$), and so on.
The first realization in RSE thus contains a mixture 
of the signal drawn from all the $50$ realizations in SE.  For the
second realization in RSE, the signal for all the modes in ${\mathcal
  A}_1$ is drawn from [SE]$_2$, and the signal for all the modes in
${\mathcal A}_2$ is drawn from [SE]$_3$ and so on. The second 
realization in RSE also contains signal drawn from all the $50$ 
realizations in SE. Further, there is no signal which is common 
between the first and second realization in RSE. The $50$
realizations in RSE have all been constructed in this fashion such
that each realization of RSE contains a mixture of the signal from all
the $50$ realizations in SE. Further, none of the realizations 
in RSE have any signal in common. The right-hand panel of
Fig. \ref{fig:HI_map} shows a section through one realization of the
RSE for $\xb=0.5$.

We do not expect the 
signal in the  modes drawn from SE$_1$ to be correlated with those
drawn from SE$_2$, etc.  
We therefore expect the average  trispectrum
$\bar{T} (k_i,k_j)$ to be at least $50$ times smaller for  RSE
as compared to SE.  For the purpose of this work we have assumed that 
$\bar{T} (k_i,k_j) \approx 0$ for RSE. 
Further,
since the entire signal in SE is also present in RSE, we expect 
$\bar{P} (k_i)$ and  $\overline{P^2}(k_i)$  to have {\em exactly} 
the same value in both SE and RSE. The RSE, therefore, provides  an 
independent estimates of  $\overline{P^2}(k_i)$. We have used this  
to estimate  the values which the diagonal elements  of $\cov_{ij}$ 
(equation \ref{eq:cov}) are expected  to have if the EoR signal were  a
 Gaussian random field with $\bar{T} (k_i,k_j) = 0$. 
It thus becomes  possible to interpret  any deviations from this 
as arising from  $\bar{T} (k_i,k_j)$ due to 
the non-Gaussianity in the EoR 21-cm signal. 

\subsection{Ensemble of Gaussian Random Ensembles}
\label{sec:egre}
The second complication arises from the fact that the SE has a 
finite number of realizations. To appreciate this we
construct the Gaussian Random Ensemble (GRE). The GRE, like the SE,
contains $50$ realizations of the 21-cm signal, the signal in each
realization however is a Gaussian random field. The signal at any
mode $\k$ in the $i$~th bin is calculated using
\begin{equation}
\tTb ({\k}) = \sqrt{\frac{ V \bar{P} ({k_i})}{2}} [a({\k})+ i b({\k})]
\end{equation}
where $a({\k})$ and $b({\k})$ are two real valued independent Gaussian
random variables of unit variance, and $\bar{P} ({k_i})$ is
the bin-averaged power spectrum calculated from SE.  The middle panel
of Fig. \ref{fig:HI_map} shows a section through one realization of
the GRE for $\xb=0.5$.

The bin-averaged power spectrum estimated from any single realization
in GRE will be different from $\bar{P} ({k_i})$. Further, the
bin averaged power spectrum estimated using all $50$ members of GRE,
which we refer to as $[\bar{P} ({k_i})]_G$, will also differ
from $\bar{P} ({k_i})$ because of the limited number of
realizations.  Similarly, the off -diagonal terms of the
error-covariance $[\cov_{ij}]_G$ estimated from GRE will not be zero
but will have random fluctuations around zero due to the limited
number of realizations.  It is thus necessary to compare the
$\cov_{ij}$ estimated from SE against the random fluctuation of
$[\cov_{ij}]_G$ in order to determine whether $\cov_{ij}$ estimated
from SE is statistically significant or not. The issue now is to
estimate the variance of the covariance $[\cov_{ij}]_G$.  We have used
$50$ independent GREs to construct an Ensemble of Gaussian Random
Ensembles (EGRE) which we have used to estimate the variance 
$[\delta \cov_{ij}]_G^2$ of $[\cov_{ij}]_G$.  In summary, we cannot
straightaway interpret the non-zero off-diagonal terms in $\cov_{ij}$
as arising from non-Gaussianity in the EoR 21-cm signal. It is
necessary to assess the statistical significance of the non-zero
values by comparing them against $[\delta \cov_{ij}]_G$ estimated from
the EGRE.

\section{Results}
\label{sec:results}
Fig. \ref{fig:HI_map} shows three 21-cm maps corresponding to
individual realizations drawn from the SE, GRE and RSE respectively.
The simulations all correspond to the same neutral fraction $\xb=0.5$
and they all have the same bin averaged power spectrum $\bar{P}
(k_i)$. It is believed that at the length scales which will be 
probed by observations the EoR 21-cm signal (in terms of both power 
spectrum and variance) peaks around $\xb \approx 0.5$ (see e.g. 
\citealt{mcquinn07,lidz08,barkana09,choudhury09b,mesinger11,jensen13,majumdar13,iliev13,patil14,watkinson14,majumdar15}), 
and we have thus restricted the entire discussion of this section to 
the situation where $\xb=0.5$.  At this stage we expect a little
less than $50 \%$ of the volume to be occupied by ionized bubbles. 
These bubbles, which are quite distinctly visible in the left-hand
panel, cause the EoR 21-cm signal to be significantly non-Gaussian at
$\xb=0.5$. This is quite apparent if we compare the EoR signal to the
central panel which is a Gaussian random field. There are no bubbles
visible in the central panel. The Randomized Signal (shown in the
right most panel of the same figure), which has a much smaller
trispectrum compared to the EoR signal, looks quite distinct from both
the other cases.

\begin{figure}
\psfrag{pk}[c][c][1][0]{\large ${\Delta^2_{\rm b}}\, \, ({\rm mK^2})$}
\psfrag{k}[c][c][1][0]{\large $k\, \, ({\rm Mpc}^{-1}$)}
\psfrag{10}[c][c][1][0]{\large $10$}
\psfrag{150mpc}[c][c][1][0]{\large $150{\rm Mpc}$}
\psfrag{215mpc}[c][c][1][0]{\large $215{\rm Mpc}$}
\centering
\includegraphics[width=.33\textwidth, angle=-90]{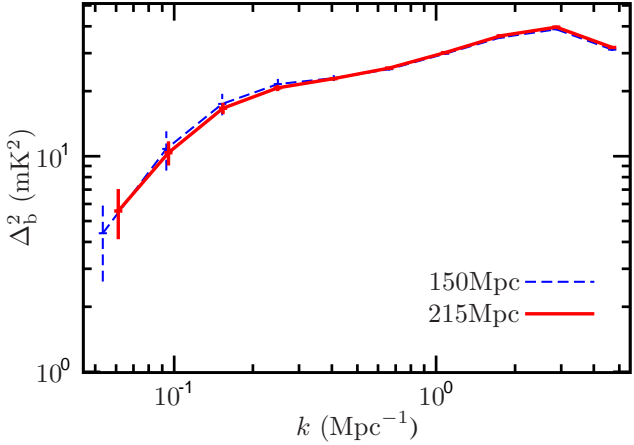}
\caption{The dimensionless brightness temperature power spectrum
  $\Delta_{\rm b}^2(k)$ and its $1\sigma$ error bars for $\xb=0.5$.
  The results are shown for simulations with the two different box
  size of $150{\rm Mpc}$ and $215{\rm Mpc}$, respectively.}
\label{fig:pk_HI}
\end{figure}

Fig. \ref{fig:pk_HI} shows the mean squared brightness temperature
fluctuation of the EoR 21-cm signal ${\Delta_{\rm b}^2}(k)=k^3
\bar{P}(k)/(2{\rm \pi})^2$ as a function of $k$.  This essentially
is a measure of the bin averaged 21-cm power spectrum $\bar{P} (k)$
estimated from SE. The $k$ range $ k_{\rm min} = 2.09 \times 10^{-2}
\, {\rm Mpc}^{-1}$ to $k_{\rm max} = 5.61 \, {\rm Mpc}^{-1}$ has been
divided in 10 equally spaced logarithmic bins with $\Delta k_i/k_i
\approx 0.48$.  We have maintained the same bin widths for both the
simulation volumes $V_1=[150\, {\rm Mpc}]^3$ and $V_2=[215\, 
{\rm Mpc}]^3$.  However, we notice that the value of $k_i$, the average
$k$ value corresponding to a particular bin, varies from $V_1$ to
$V_2$ (Fig. \ref{fig:pk_HI}).  This variation arises because the
exact number and values of the Fourier modes in a particular bin
changes if we change the simulation volume even though $\Delta k_i$ is
fixed. Comparing the results from the two simulation volumes, we see
that there is very little change in the power spectrum between $V_1$
and $V_2$.  This indicates that the simulation volumes used here are
sufficiently large so that the power spectrum has converged. The error
bars shown in the figure correspond to the $1-\sigma$ error $\delta
P(k_i)=\sqrt{\cov_{ii}}$ estimated from SE. We notice that the error
bars change from $V_1$ to $V_2$, the errors being smaller for the
larger simulation. This arises from the $\cov_{ii} \propto 1/V$
dependence (eq. \ref{eq:cov}) discussed earlier. A detailed analysis
of the covariance matrix $\cov_{ij}$ follows.

\begin{figure}
\psfrag{cov}[c][c][1][0]{\large $\cov_{ii}  \, \, ({\rm Mpc}^6\,{\rm mK}^4)$}
\psfrag{k}[c][c][1][0]{\large $k\, \, ({\rm Mpc}^{-1})$}
\psfrag{10}[c][c][1][0]{\large $10$}
\psfrag{scaled}[c][c][1][0]{\large Scaled}
\centering
\includegraphics[width=.33\textwidth, angle=-90]{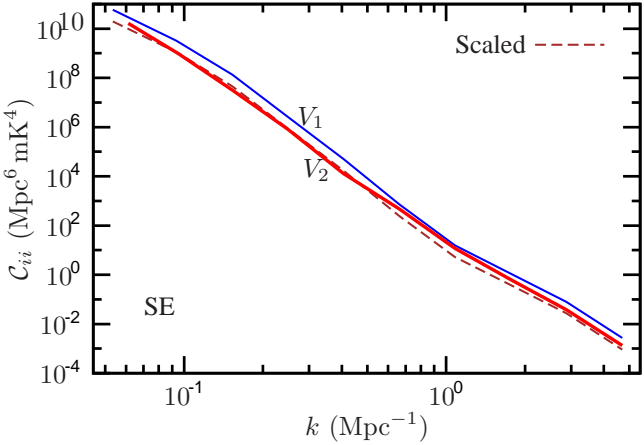}
\put(-191, -121){\large{SE}}
\put(-133, -49){\large $V_1$}
\put(-131, -69){\large $V_2$}
\caption{This shows $\cov_{ii}$ for SE considering both the 
simulation volumes $V_1$ and $V_2$. We  also show 
 $(V_1/V_2) \, [\cov_{ii}]_{V_1}$ where $\cov_{ii}$
determined for $V_1$ has been {\bf scaled} to account for 
the $1/V$ dependence predicted by equation (\ref{eq:cov}). }
\label{fig:cov_SE}
\end{figure}

We now shift our attention to the error covariance matrix $\cov_{ij}$
which is the main focus of this paper. Fig. \ref{fig:cov_SE} shows
the diagonal elements $\cov_{ii}$ as a function of $k$ for the two
different simulation volumes $V_1$ and $V_2$. We have also shown
$(V_1/V_2) \, [\cov_{ii}]_{V_1}$ where the matrix elements determined
for $V_1$ have been {\em scaled} to account for the $1/V$ dependence
predicted by equation (\ref{eq:cov}).  We see that the scaled elements are
in reasonable agreement with the results for $V_2$, roughly indicating
that the error-covariance has converged within the simulation volume
which we have used here. We see that the values of the covariance
matrix span a very large dynamical range, and it is not very
convenient to analyse this if we are looking for relatively small
changes in the values. We find that it is much more convenient to
instead use the {\em dimensionless} covariance matrix $\dcov_{ij}$
which  is   defined as
\begin{equation}
\dcov_{ij}=\frac{\cov_{ij} \, V \ k_i^{3/2} k_j^{3/2}}{(2 \pi)^2
 \bar{P}(k_i) \,  \bar{P}(k_j)} \, ,
\label{eq:dcov}
\end{equation}
and which, using equation (\ref{eq:cov}), can be expressed as 
\begin{equation}
\dcov_{ij}=A_i^2 \left(
\frac{k_i}{\Delta k_i} \right) \delta_{ij}+ t_{ij}
\label{eq:dcov1}
\end{equation}
where 
\begin{equation}
t_{ij}=\frac{\bar{T} (k_i,k_j) \ k_i^{3/2} \, k_j^{3/2}  }{(2
  \pi)^2  \bar{P} (k_i) \,  \bar{P} (k_j)} \, ,
\label{eq:dtrisp}
\end{equation}
is the dimensionless bin-averaged trispectrum 
and 
\begin{equation}
A_i=\sqrt{\frac{\overline{P^2}(k_i)}{[\bar{P}(k_i)]^2}} \,.
\label{eq:Ai}
\end{equation}
is a number of order unity introduced in \citet{mondal15}. The value
of $A_i$ is expected to vary from bin to bin. We also expect its value
to vary if we change the simulation volume. However, all these
variations are expected to be small, and we may expect a value $A_i
\approx 1$ in most situations.

\begin{figure*}
\psfrag{cov}[c][c][1][0]{\large $\dcov_{ii}$}
\psfrag{k}[c][c][1][0]{\large $k\, \, ({\rm Mpc}^{-1}$)}
\psfrag{10}[c][c][1][0]{\large $10$}
\psfrag{150mpc}[c][c][1][0]{\large $150{\rm Mpc}$}
\psfrag{215mpc}[c][c][1][0]{\large $215{\rm Mpc}$}
\psfrag{prediction}[c][c][1][0]{\large Prediction\,}
\centering
\includegraphics[width=.35\textwidth, angle=-90]{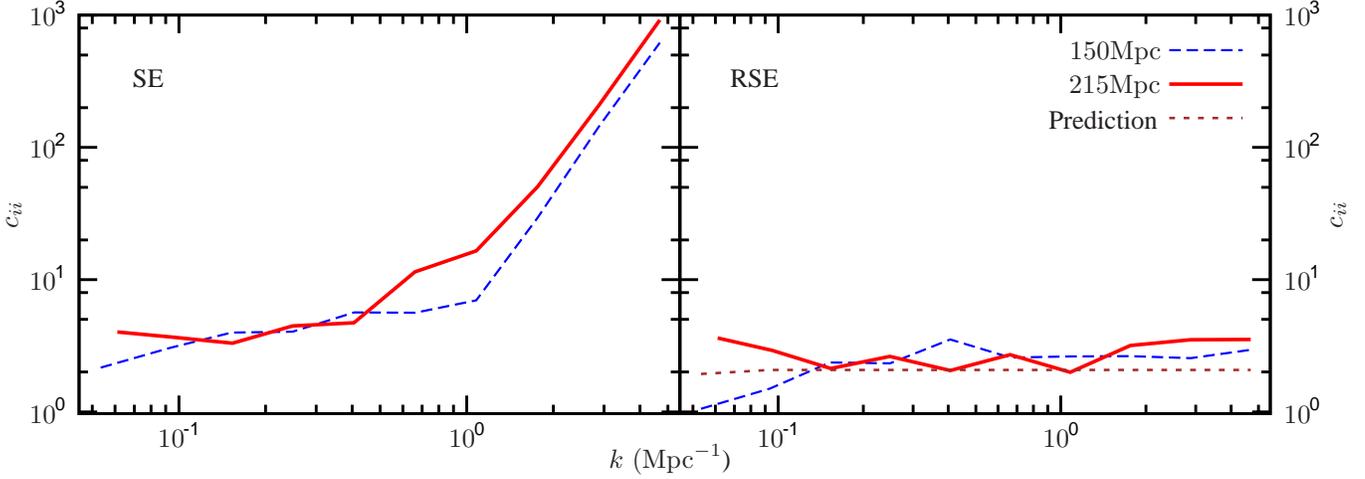}
\put(-457, -31){\large{SE}}
\put(-231, -31){\large{RSE}}
\caption{This shows  $\dcov_{ii}$ for SE (left) and RSE (right). The 
{\bf prediction} based on using the constant $A$ from \citet{mondal15}
in equation (\ref{eq:dcovrse}) is also shown in the right-hand panel.}
\label{fig:cov_SE_RSE}
\end{figure*}

The left-hand panel of Fig. \ref{fig:cov_SE_RSE} shows $\dcov_{ii}$, the
diagonal elements of the dimensionless covariance matrix, as a
function of $k$. The volume dependence of $\cov_{ii}$ has been scaled
out in the definition of $\dcov_{ii}$ (equation \ref{eq:dcov}), and we do
not expect the $\dcov_{ii}$ values to change if we vary the volume
provided that the error-covariance has converged within the simulation
volume.  We find that the values of $\dcov_{ii}$ obtained from the two
different volumes $V_1$ and $V_2$ are consistent with each other over
the range $0.1 \le k \le 0.5 \, {\rm Mpc}^{-1}$. The values obtained
from $V_2$, however, are $\sim 1.5$ times larger than those obtained
from $V_1$ at larger values $k > 0.5 \, {\rm Mpc}^{-1}$. The
difference at small $k$ ($k < 0.1 \, {\rm Mpc}^{-1}$) may be
attributed to the cosmic variance of the error-covariance and is
possibly not statistically significant. However, the differences
between $V_1$ and $V_2$ at large $k$ appears to be significant. We
find that the smaller volume $V_1$ is under-estimating the
error-covariance relative to $V_2$, indicating that for $k > 0.5 \,
{\rm Mpc}^{-1}$ the error-covariance has not converged within the
simulation volume.  One would naively expect convergence issues to be
more important at large scales which are comparable to the simulation
size.  The fact that the error-covariance appears to have converged at
large scales ($0.1 \le k \le 0.5 \, {\rm Mpc}^{-1}$) while it seems to
have not converged at small scales ($k > 0.5 \, {\rm Mpc}^{-1}$) is
quite counter intuitive, and we shall address this a little later.
  
The right-hand panel of Fig. \ref{fig:cov_SE_RSE} shows $\dcov_{ii}$
estimated from the RSE for which we expect $t_{ij} \approx 0$, whereby
\begin{equation}
[\dcov_{ii}]_{\rm RSE}=A_i^2 \left(
\frac{k_i}{\Delta k_i} \right) \,.
\label{eq:dcovrse}
\end{equation}
This gives an estimate of the error-covariance that would be expected
if the EoR signal were a Gaussian random field. As expected, we see
that the values of $[\dcov_{ii}]_{\rm RSE}$ are below those estimated
from SE.  \citet{mondal15} have estimated the value of $A_i$ in a
completely independent manner by fitting the behaviour of the 
${\rm SNR}$ as a function of $N_k$. The latter method ignores the fact
that $A_i$ varies from bin to bin, and returns just a single value of
$A$ which is $A=0.98$ for $\xb=0.5$.  We have also plotted
$[\dcov_{ii}]_{\rm RSE}$ using this $A$ value and the ${k_i}/{\Delta 
k_i}$ values corresponding to the $k$ bins in $V_1$. We note that
${\Delta k_i}/{k_i}  \approx 0.48$, though the actual value changes
somewhat from bin to bin.  We find that $[\dcov_{ii}]_{\rm RSE}$
estimated from the $V_1$ and $V_2$ RSE simulations, and also from
equation (\ref{eq:dcovrse}) using the constant $A$, are all consistent with
one another.This consistency, in a sense, also validates the idea that
the RSE allows us to independently estimate the error-covariance that
would be expected if the EoR signal were a Gaussian random field
(equation \ref{eq:dcovrse}).

\begin{figure}
\psfrag{cov}[c][c][1][0]{\large $\dcov_{ii}$}
\psfrag{k}[c][c][1][0]{\large $k\, \, ({\rm Mpc}^{-1}$)}
\psfrag{10}[c][c][1][0]{\large $10$}
\psfrag{SE}[c][c][1][0]{\large $150{\rm Mpc}$: SE\, \, \, \, \, \, \, \, \,}
\psfrag{RSE}[c][c][1][0]{\large RSE}
\psfrag{RSE25}[c][c][1][0]{\large RSE25}
\psfrag{RSE10}[c][c][1][0]{\large RSE10}
\centering
\includegraphics[width=.33\textwidth, angle=-90]{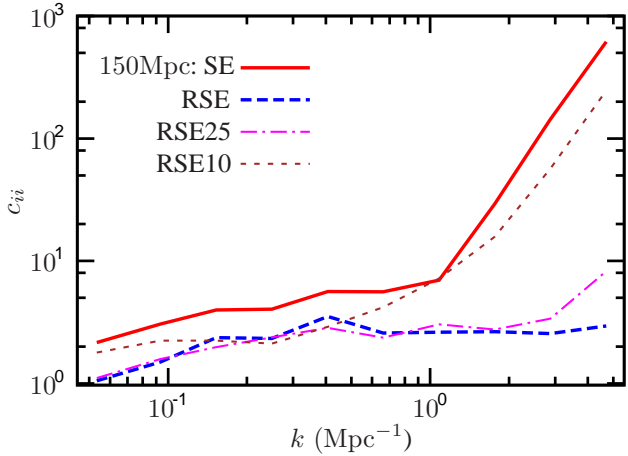}
\caption{This shows $\dcov_{ii}$ as determined from  SE, RSE, RSE25 and RSE10.}
\label{fig:cov_SE_RSE_150}
\end{figure}

We further illustrate the idea behind the RSE and also validate this
in Fig. \ref{fig:cov_SE_RSE_150}. Recollect that each realization in
the RSE contains a mixture of signal from $50$ independent
realizations of the EoR signal, and we expect $t_{ij}$ for RSE to be
at least $50$ times smaller than $t_{ij}$ estimated from SE.  In
addition to RSE, we also show results for RSE10 and RSE25. Each
realization in RSE10 has signal drawn from $10$ independent
realization from SE instead of $50$. We expect $t_{ij}$ for RSE10 and
RSE25 to be respectively around $10$ and $25$ times smaller than
$t_{ij}$ estimated from SE.  Starting from SE (equation \ref{eq:dcov1}) ,
we expect the values of $\dcov_{ii}$ to slowly approach
equation (\ref{eq:dcovrse}) as we move from RSE10 to RSE25 and then to RSE.
This transition is clearly seen in Fig. \ref{fig:cov_SE_RSE_150}.
There is very little change in the values of $\dcov_{ii}$ from RSE25
to RSE50 ( except possibly at the largest $k$ value). This validates
the assumption that $t_{ij} \approx 0$ for the RSE.

\begin{figure}
\psfrag{cov}[c][c][1][0]{\large $t_{ii}$}
\psfrag{k}[c][c][1][0]{\large $k\, \, ({\rm Mpc}^{-1}$)}
\psfrag{10}[c][c][1][0]{\large $10$}
\psfrag{150mpc}[c][c][1][0]{\large $150{\rm Mpc}$}
\psfrag{215mpc}[c][c][1][0]{\large $215{\rm Mpc}$}
\centering
\includegraphics[width=.33\textwidth, angle=-90]{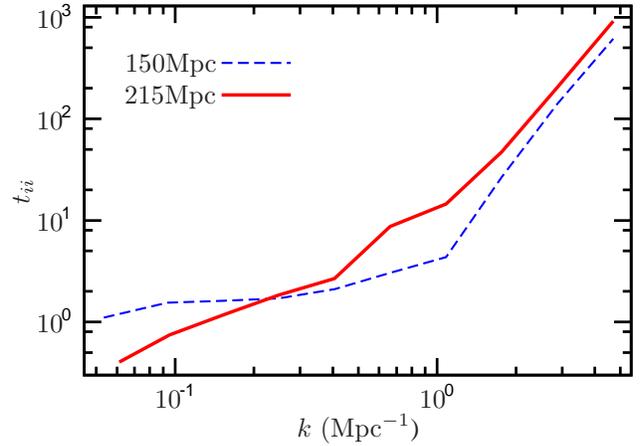}
\caption{This shows $t_{ii}$ estimated from the two different 
simulation volumes $V_1$ and $V_2$.}
\label{fig:trispec}
\end{figure}

The difference $\cov_{ii}-[\cov_{ii}]_{\rm RSE}$ gives an estimate of
the bin-averaged trispectrum. Here we have used
$t_{ii}=\dcov_{ii}-[\dcov_{ii}]_{\rm RSE}$ to estimate the
dimensionless bin-averaged trispectrum for which the results are shown
in Fig. \ref{fig:trispec}. We see that the results for the two
different simulation volumes look quite similar, though there are some
differences in the actual values. The $t_{ii}$ values estimated from
the larger volume $V_2$ are larger than those estimated from $V_1$ at
$k > 0.2 \, {\rm Mpc}^{-1}$.  The $t_{ii}$ values differ by a nearly
constant ratio of $1.5$ at $k > 1 \, {\rm Mpc}^{-1}$. The trend is
reversed at $k < 0.2 \, {\rm Mpc}^{-1}$ where the values estimated
from $V_1$ are larger than those from $V_2$. Taken at face value,
these discrepancies in the values of $t_{ii}$ between the two
different simulation volumes indicate that the trispectrum has not
converged within the simulation volume. We note, however, that it is
necessary to be cautious before arriving at such a conclusion because
we have no estimate of the cosmic variance for $t_{ii}$. The
discrepancy at large $k$ is possibly genuine, whereas the discrepancy
at small $k$ is possibly influenced by the cosmic variance.  For the
subsequent discussion in this paper we focus on the larger volume
$V_2$ assuming that the results are representative of what would be
expected for an even larger volume.

\begin{figure}
\psfrag{covratio}[c][c][1][0]{\large $t_{ii}/[\dcov_{ii}]_{\rm RSE}$}
\psfrag{k}[c][c][1][0]{\large $k\, \, ({\rm Mpc}^{-1}$)}
\psfrag{10}[c][c][1][0]{\large $10$}
\psfrag{150mpc}[c][c][1][0]{\large $150{\rm Mpc}$}
\psfrag{215mpc}[c][c][1][0]{\large $215{\rm Mpc}$}
\centering
\includegraphics[width=.33\textwidth, angle=-90]{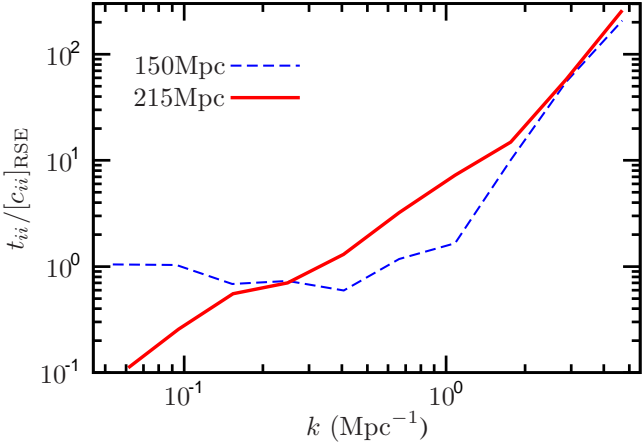}
\caption{This shows the ratio $t_{ii}/[\dcov_{ii}]_{\rm RSE}$ estimated from 
the two different  simulation volumes $V_1$ and $V_2$.  }
\label{fig:ratio}
\end{figure}

We see (Fig. \ref{fig:trispec}) that we have $t_{ii} \sim 1$ for $k
\sim 0.1 \, {\rm Mpc}^{-1}$, and it increases quite rapidly with
$t_{ii} \sim 10$ and $\sim 10^3$ at $k \sim 1 \, {\rm Mpc}^{-1}$ and $
\sim 5 \, {\rm Mpc}^{-1}$ respectively. In contrast, we have
$[\dcov_{ii}]_{\rm RSE} \sim 5$ for nearly the entire $k$ range
(Fig. \ref{fig:cov_SE_RSE}).  We thus expect the error-covariance
$c_{ii}$ to be largely dominated by the trispectrum $t_{ii}$ for
nearly the entire $k$ range that we have considered here. Fig.
\ref{fig:ratio} shows the ratio $t_{ii}/[\dcov_{ii}]_{\rm RSE}$ which
quantifies the relative magnitudes of the two terms that contribute to
$c_{ii}$ (equation \ref{eq:dcov1}).  We see that the two terms make roughly
equal contributions in the range $0.2 \le k \le 0.3 \, {\rm
  Mpc}^{-1}$. The relative contribution from the trispectrum increases
quite steeply with increasing $k$.  At the largest $k$ value ($ \sim 5
\, {\rm Mpc}^{-1}$), the contribution from the trispectrum is $\sim
200$ times larger than the error-covariance that we would expect if
the EoR signal were a Gaussian random field.

\begin{figure}
\psfrag{k}[c][c][1][0]{\large $k\, \, ({\rm Mpc}^{-1})$}
\psfrag{cov}[c][c][1][0]{\large $r_{ij}$}
\centering
\includegraphics[width=.33\textwidth, angle=-90]{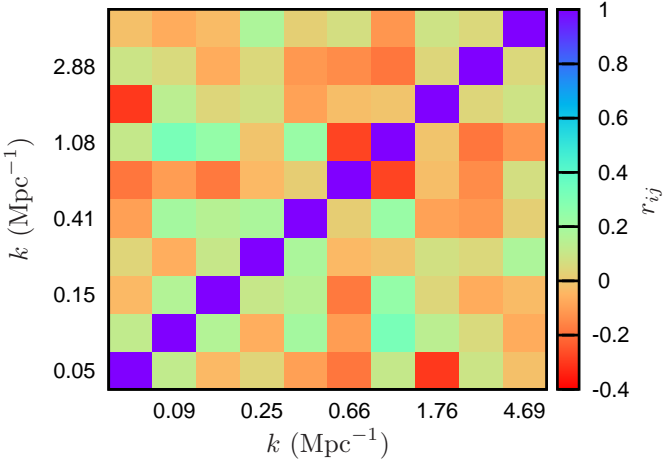}
\caption{This shows $r_{ij}$ estimated for a GRE.}
\label{fig:cov_rand}
\end{figure}
\begin{figure*}
\psfrag{k}[c][c][1][0]{\large $k\, \, ({\rm Mpc}^{-1})$}
\psfrag{cov}[c][c][1][0]{\large $r_{ij}$}
\centering
\includegraphics[width=.345\textwidth,
  angle=-90]{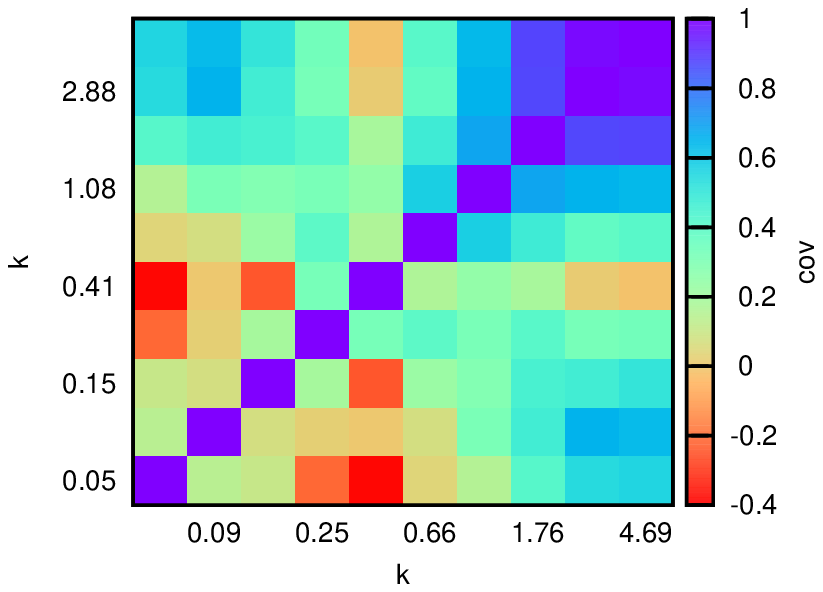}
\includegraphics[width=.345\textwidth,
  angle=-90]{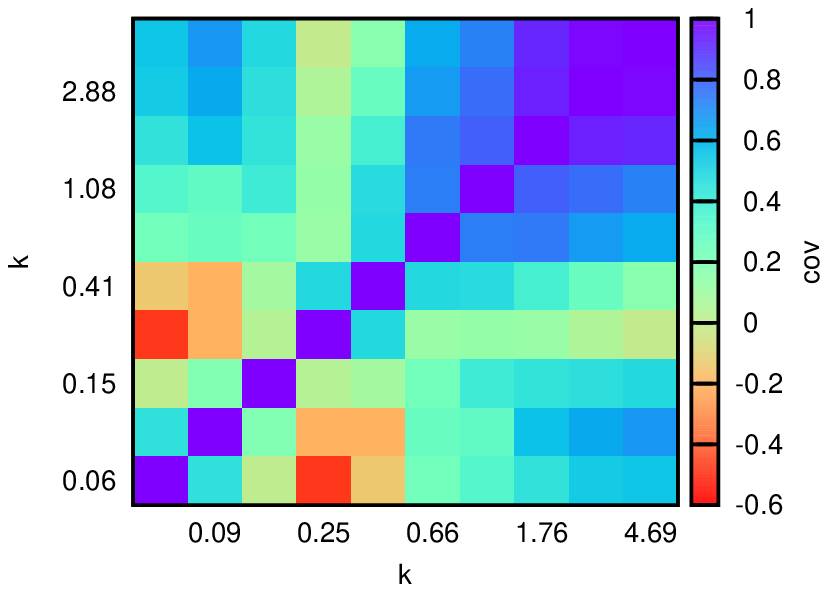}
\caption{This shows $r_{ij}$ estimated for SE considering both the 
simulation volumes $V_1$ (left) and $V_2$ (right).}
\label{fig:cov_sig}
\end{figure*}

We now shift our focus to the off-diagonal elements of $\dcov_{ij}$  which 
quantify the correlation between  the errors at different $k$ bins. 
Since the diagonal terms  $\dcov_{ii}$ span a pretty large dynamical 
range, it is more convenient to consider the correlation coefficient
\begin{equation}
r_{ij} = \frac{\dcov_{ij}}{\sqrt{\dcov_{ii} \, \dcov_{jj}}} \, 
\label{eq:ccof}
\end{equation}
 instead of directly analysing the off-diagonal terms of $\dcov_{ij}$.
 The values of $r_{ij}$ are, by definition, restricted to lie in the
 range $-1 \le r_{ij} \le 1$, the values $r_{ij}=1$ and $-1$
 indicating that the errors in the $i$ and $j$ bin are fully
 correlated and anti-correlated respectively. Intermediate values ($-1
 < r_{ij} < 1$) indicate partial correlation or anti-correlation, and
 $r_{ij}=0$ indicates that the errors in the $i$ and $j$ bins are
 uncorrelated. Also note that we have $r_{ij}=1$ for all the diagonal
 elements of $r_{ij}$.  We first consider the GRE for which the errors
 in the different bins are uncorrelated. Fig. \ref{fig:cov_rand}
 shows $r_{ij}$ estimated using a single GRE. We see that in addition
 to the diagonal elements which have value $r_{ii}=1$, the
 off-diagonal elements also have non-zero values. As discussed in
 Section \ref{sec:egre}, these non-zero values are from random
 fluctuations which are a consequence of the limited number of
 realizations in the GRE. Fig. \ref{fig:cov_sig} shows $r_{ij}$
 estimated from SE. We see that the results from both the simulation
 volumes of SE look very similar. Comparing the SE with the GRE, we
 see that while the $r_{ij}$ values in Fig. \ref{fig:cov_rand} (GRE)
 appear to be quite random, Fig. \ref{fig:cov_sig} (SE) exhibits
 some sort of an organized pattern. The most prominent feature which
 we notice is that the errors in the five largest $k$ bins ($k > 0.5 \,
 {\rm Mpc}^{-1}$) are strongly correlated. Further, the errors in the
 three smallest $k$ bins ($k < 0.3 \, {\rm Mpc}^{-1}$) are correlated
 with the three largest $k$ bins ($k > 1 \, {\rm Mpc}^{-1}$).
 Finally, we also find a relatively weak anti-correlation between the
 two smallest $k$ bins ($k < 0.1\, {\rm Mpc}^{-1}$) and the
 intermediate bins $\sim 0.2\, - \, 0.4 \, {\rm Mpc}^{-1}$.

\begin{figure*}
\psfrag{k}[c][c][1][0]{\large $k\, \, ({\rm Mpc}^{-1})$}
\psfrag{cov}[c][c][1][0]{\large $r_{ij}$}
\psfrag{RSE}[c][c][1][0]{\large RSE}
\psfrag{150mpc}[c][c][1][0]{\large $150{\rm Mpc}$}
\psfrag{215mpc}[c][c][1][0]{\large $215{\rm Mpc}$}
\centering
\includegraphics[width=1.31\textwidth, angle=-90]{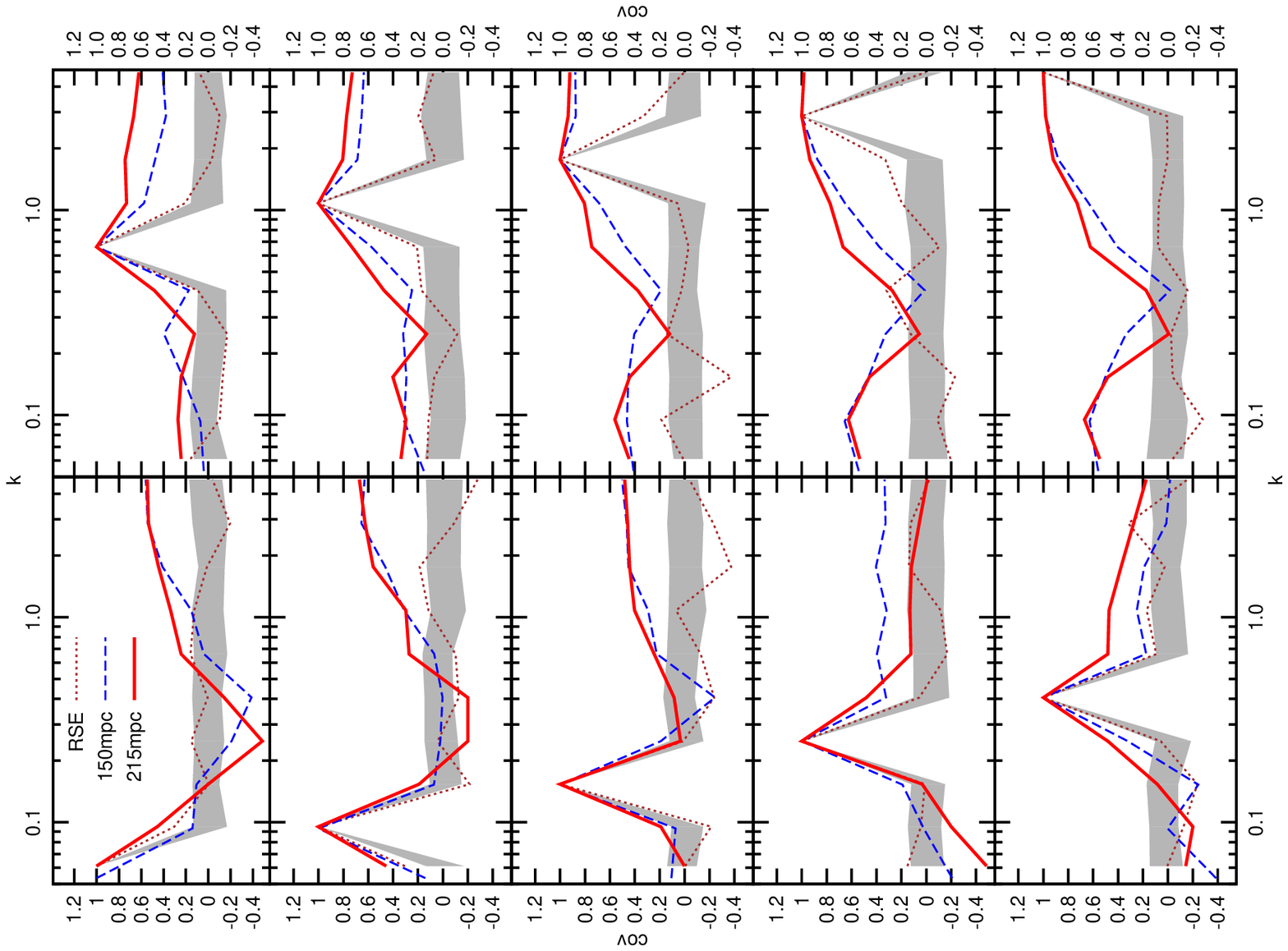}
\caption{This shows $r_{ij}$ estimated for SE considering both the
  simulation volumes $V_2$ (solid) and $V_1$ (dashed). We also show
  $r_{ij}$ estimated from RSE (dotted) with $V_2$. The shaded region
  represents the $[\delta r_{ij}]_G$ which quantifies the fluctuation
  of the off-diagonal terms around $[r_{ij}]_G=0$ expected for a
  Gaussian random field.}
\label{fig:all_cov}
\end{figure*}

Fig. \ref{fig:all_cov} shows the $r_{ij}$ values estimated from SE
for both the simulation volumes $V_1$ and $V_2$.  Each panel of the
figure corresponds to a fixed value of $i$, and it shows $r_{ij}$ as a
function of $k_j$.  We have used the EGRE (Section \ref{sec:egre}) to
estimate $[\delta r_{ij}]_G$ which quantifies the fluctuation of the
off-diagonal terms around $[r_{ij}]_G=0$ expected for a Gaussian
random field.  For reference, we have also shown $r_{ij}$ estimated
from RSE with $V_2$.  Note that in all cases we have $r_{ij}=1$ for
the diagonal terms which have $j=i$.

We expect $[t_{ij}]_{\rm RSE} \approx 0$, which implies that we also
expect $[r_{ij}]_{\rm RSE}=0$ for the off-diagonal terms.  We find
that the values estimated from RSE are nearly always within the shaded
region corresponding to $[\delta r_{ij}]_G$, indicating that our
results are indeed consistent with $[r_{ij}]_{\rm RSE}=0$. This is yet
another validation of the fact that the method by which we have
generated the RSE actually destroys the correlation between the signal
at different Fourier modes and results in $[t_{ij}]_{\rm RSE} \approx
0$.  The results from $V_1$ and $V_2$ are quite similar for
SE. Further, there are several regions where the $r_{ij}$ values for
SE are outside the shaded region.  We interpret these as being
statistically significant and discuss these below.  We find that the
errors in the five largest bins ($k > 0.5 \, {\rm Mpc}$) are strongly
correlated with the correlation coefficient having values $r_{ij} \ge
0.6$. The correlation increases to $r_{ij} \ge 0.9$ if we consider
just the three largest $k$ bins.  The errors in the three smallest $k$
bins ($k < 0.2 \, {\rm Mpc}$) are also correlated with the errors in
the five largest $k$ bins. The errors in the two smallest $k$ bins ($k
< 0.1 \, {\rm Mpc}$), however, are weakly anti-correlated with the
errors in the $4$-th and $5$-th bins ($0.2 < k < 0.4 \, {\rm Mpc}$).

\section{Summary and Discussion}
\label{sec:summary}
The error-covariance matrix of the EoR 21-cm power spectrum is an
important ingredient for making predictions for ongoing and future
experiments to detect the EoR signal. In this work we only consider
the errors which are intrinsic to the EoR 21-cm signal, i.e. the
cosmic variance, and ignore the system noise arising from
radio-interferometric observations. The EoR 21-cm signal becomes
increasingly non-Gaussian as reionization proceeds.  Non-Gaussianity
introduces correlations between the signal in different Fourier modes,
this being quantified through the bispectrum, trispectrum, etc.  While
the power spectrum itself does not tell anything as to whether the
underlying signal is Gaussian or non-Gaussian, we show that the
error-covariance matrix $\cov_{ij}$ for the binned power spectrum is
sensitive to the non-Gaussianity through the bin averaged trispectrum
$ \bar{T} (k_{i},k_{j})$ which appears in equation (\ref{eq:cov}).

The error covariance matrix scales inversely with the volume as
$\cov_{ij}\propto V^{-1}$, and it is more convenient to analyse the
dimensionless error covariance matrix $\dcov_{ij}$ (equation \eqref{eq:dcov})
which is independent of volume. We have used an ensemble of $50$
independent realizations of the simulated EoR 21-cm signal (referred
to as the SE) to estimate $\dcov_{ij}$.  The
entire analysis was restricted to a single neutral fraction $\xb=0.5$.
The left-hand panel of Fig. \ref{fig:cov_SE_RSE} shows $\dcov_{ii}$, the
diagonal elements of $\dcov_{ij}$, as a function of $k$.  We can
interpret each diagonal element $\dcov_{ii}$ as the dimensionless
error variance for the power spectrum estimated in the corresponding
bin. For the $\Delta k_i$ bins used here, we expect the dimensionless 
error variance to have a value  $\dcov_{ii} \approx 2$ across all 
the $k$ bins if the EoR signal is a Gaussian random field. 
We find a roughly constant value $\dcov_{ii} \sim 5$ in the $k$
range $0.05 \le k \le 0.5 \, {\rm Mpc}^{-1}$, the value of
$\dcov_{ii}$ increases sharply beyond $k \ge 0.5 \, {\rm Mpc}^{-1}$
and we have $\dcov_{ii} \sim 10^3$ at $k \sim 5.0 \, {\rm Mpc}^{-1}$.
We see that the actual error in the estimated EoR 21-cm power spectrum is 
considerably in excess of the error predicted for a Gaussian random field.  
This discrepancy arises because the EoR \HI distribution is dominated 
by several large ionized bubbles (left-hand panel of Fig. \ref{fig:HI_map}) 
and  the emanating 21-cm signal is not a Gaussian random field.

The diagonal elements $\dcov_{ii}$ are the sum of two parts
(equation \ref{eq:dcov1}).  The first part $A_i^2 \left( {k_i}/{\Delta k_i}
\right)$ is the contribution that would arise if the EoR signal were a
Gaussian random field. In this case it is possible to reduce the error
covariance $\dcov_{ii}$ by increasing the bin width or equivalently
combining a larger number of independent Fourier modes.
Non-Gaussianity, however, introduces an extra term $t_{ii}$ which is
the dimensionless bin averaged trispectrum.  As a consequence the
dimensionless error variance $\dcov_{ii}$ does not decrease beyond a
limiting value, and it is not possible to decrease the error beyond
this by increasing the number of Fourier modes in the bin.

The SE provides an estimate of the total
dimensionless error variance $\dcov_{ii}$, however it is not possible to
separately estimate the two parts $A_i^2 \left( {k_i}/{\Delta k_i}
\right)$ and $t_{ii}$ using SE.  We have overcome this problem by
constructing the RSE in which each
realization contains a mixture of the signal from all realizations of
SE. This destroys the correlation between the signal at different
Fourier modes, and we have $t_{ii} \approx 0$. Since the entire signal
in SE is also present in RSE, the RSE provides an independent estimate
of the $\dcov_{ii}$ that would be expected if the EoR 21-cm signal
were a Gaussian random field (i.e. $[\dcov_{ii}]_{\rm RSE}=A_i^2
\left({k_i}/{\Delta k_i} \right)$).  The right-hand panel of Fig.
\ref{fig:cov_SE_RSE} shows $[\dcov_{ii}]_{\rm RSE}$ as a function of
$k$. We find that the $[\dcov_{ii}]_{\rm RSE}$ show little variation
with $k$ with values in the range $2 \le [\dcov_{ii}]_{\rm RSE} \le
5$.  This is consistent with what we expect from $A_i \approx 1$ and
${\Delta k_i}/{k_i} \approx 0.48$, note that the actual values of
$A_i$ and $\bar{P}(k_i)/{k_i}$ vary from bin to bin.

The difference $\dcov_{ii}-[\dcov_{ii}]_{\rm RSE}$ gives an estimate
of the dimensionless bin-averaged trispectrum $t_{ii}$. We find
(Fig. \ref{fig:trispec}) that the value of $t_{ii}$ increases
monotonically with $k$. We have $t_{ii} \sim 1$ for $k \sim 0.1 \,
{\rm Mpc}^{-1}$, and it increases quite rapidly with $t_{ii} \sim 10$
and $\sim 10^3$ at $k \sim 1 \, {\rm Mpc}^{-1}$ and $ \sim 5 \, {\rm
  Mpc}^{-1}$ respectively. Fig. \ref{fig:ratio} shows the ratio
$t_{ii}/[\dcov_{ii}]_{\rm RSE}$. This quantifies the relative
magnitudes of the two terms which contribute to total error variance
$\cov_{ii}$, here $[\dcov_{ii}]_{\rm RSE}$ is the error variance that
would arise if the EoR 21-cm signal were a Gaussian random field and
$t_{ii}$ is the extra contribution to the error variance arising from
the non-Gaussianity of the EoR 21-cm signal. We find
$t_{ii}/[\dcov_{ii}]_{\rm RSE} \ge 1$ for $k \ge 0.2 \, {\rm
  Mpc}^{-1}$, the value of this ratio increases with $k$ and it is
$\sim 10$ and $\sim 200$ at $k \sim 1 \,{\rm Mpc}^{-1}$ and $k \sim 5
\,{\rm Mpc}^{-1}$ respectively. The two terms $[\dcov_{ii}]_{\rm RSE}$
and $t_{ii}$ make roughly equal contributions to $\dcov_{ii}$ in the
range $0.2 \le k \le 0.3 \, {\rm Mpc}^{-1}$. The relative contribution
from the trispectrum increases sharply at $k \ge 0.3 \, {\rm Mpc}^{-1}$ .

We find that the error variance is dominated by the trispectrum at
Fourier modes $k \ge 0.3 \, {\rm Mpc}^{-1}$. The error variance would
be severely underestimated if the EoR 21-cm signal were assumed to be
a Gaussian random field. We find that the actual error variance is
predicted to be $\sim 11$ and $\sim 200$ times larger than the
Gaussian prediction at $k \sim 1 \,{\rm Mpc}^{-1}$ and $k \sim 5
\,{\rm Mpc}^{-1}$ respectively.

We next consider the off-diagonal elements of the error covariance
$\cov_{ij}$.  The off-diagonal elements quantify the correlation
between the errors in the power spectrum estimated in different $k$
bins. The off-diagonal elements are zero for a Gaussian random field
for which the errors in the different $k$ bins are
uncorrelated. Non-Gaussianity, however, introduces correlations
between the errors at different $k$ bins (equation \ref{eq:cov}) .  We
quantify this using the dimensionless correlation coefficient $r_{ij}$
which has values in the range $-1 \le r_{ij} \le 1$, the values
$r_{ij}=1$ and $-1$ indicating that the errors in the $i$ and $j$ bin
are fully correlated and anti-correlated respectively. Intermediate
values ($-1 < r_{ij} < 1$) indicate partial correlation or
anti-correlation, and $r_{ij}=0$ indicates that the errors in the $i$
and $j$ bins are uncorrelated. We have used the SE to estimate
$r_{ij}$ for the EoR 21-cm power spectrum (Fig. \ref {fig:cov_sig}),
and the EGRE to establish the
statistical significance (Fig. \ref{fig:all_cov}).

We find that the error in the five largest $k$ bins ($k > 0.5 \, {\rm
  Mpc}^{-1}$) are strongly correlated $(r_{ij} \ge 0.6)$. We also find
a relatively weaker correlation between three smallest $k$ bins ($k < 0.3
\, {\rm Mpc}^{-1}$) and three largest $k$ bins ($k > 1 \, {\rm
  Mpc}^{-1}$).  Further, the error in the two smallest $k$ bins ($k <
0.1\, {\rm Mpc}^{-1}$) are anti-correlated with the intermediate bins
$\sim 0.2\, - \, 0.4 \, {\rm Mpc}^{-1}$. We find that this 
anti-correlation is present for both the simulation volumes
$V_1$ and $V_2$ (Fig. \ref{fig:cov_sig}) which are statistically
independent. This seems to indicate that this anti-correlation   
is a statistically significant  effect, however the origin of this
anti-correlation is not clear at 
present.

The non-linear gravitational clustering of the
  underlying density field and the presence of discrete ionized
  regions in the \HI distribution both contribute to the
  non-Gaussianity of the 21-cm signal. The non-linear gravitational
  clustering is particularly important at small scales where it leads
  to the collapse of over-dense regions to form gravitationally bound
  objects that host the luminous galaxies that subsequently reionize
  the universe. Interestingly, the over-densities are also the regions
  which get ionized first in the inside-out reionization scenario
  implemented in our simulations. Consequently, the over-dense regions
  are missing from the 21-cm signal in our simulations, and we expect
  the non-Gaussianity from the non-linear gravitational clustering to
  be subdominant to the non-Gaussianity arising from the ionized
  bubbles in the \HI distribution.  This also allows us to interpret
  the strong correlation in the error at the five largest $k$ bins ($k
  > 0.5 \, {\rm Mpc}^{-1}$).  The length-scales $(R < 13 {\rm Mpc})$
  corresponding to these Fourier modes are smaller than the size of
  the individual ionized regions (Fig. \ref{fig:HI_map}), and
  consequently the 21-cm signal in the different modes in this $k$
  range is highly correlated because it originates from the excluded
  volume of the same ionized regions. Further, the ionized regions are
  centred on the peaks of the density field which themselves are
  expected to have a clustering pattern which is related to that of
  the underlying matter distribution.  We therefore expect the ionized
  regions to be correlated with the large-scale clustering of the \HI
  distribution, a fact which is reflected in the correlation between
  the errors at large $k$ and small $k$.

This work is limited in that we have used a simple model of
reionization, and the entire analysis is restricted to a situation
where $\xb=0.5$ at $z=8$.  The predictions will be different for some
other model of reionization with different ionizing source properties,
inhomogeneous recombinations, fluctuations in the spin temperature
etc.  While the quantitative predictions are liable to change for
different reionization scenarios, this work emphasises the fact
that the non-Gaussian effects will play an important role in the error
predictions for the EoR 21-cm power spectrum. The effect of
non-Gaussianity is expected to increase further as reionization
proceeds and the neutral fraction falls below $\xb=0.5$
\citep{mondal15}.


There are several experiments like LOFAR, MWA and PAPER which are
currently underway to measure the EoR 21-cm power spectrum, and other
instruments like HERA and SKA1 LOW are expected to be functional in
future. All of these instruments target measurements of the EoR 21-cm
power spectrum in the $k$ range $0.1 \leq k \leq 2 \,{\rm Mpc^{-1}}$.
The results of this work clearly show that the the errors would
be severely underestimated under the Gaussian assumption. A proper
treatment of the error covariance matrix is crucial for correct error
predictions.  Such predictions are important to assess the prospects
of detecting the power spectrum with a particular instrument.
Further, correct error predictions are also important for interpreting
the power spectrum subsequent to a detection.  In future work we plan
to consider ongoing and future EoR experiments and carry out
comprehensive error analysis including the system noise.
\bibliographystyle{mn2e} 
\bibliography{refs}

\begin{thebibliography}{64}
\expandafter\ifx\csname natexlab\endcsname\relax\def\natexlab#1{#1}\fi

\bibitem[{{Ali}, {Bharadwaj} \& {Chengalur}(2008){Ali}, {Bharadwaj}, \&
  {Chengalur}}]{ali08}
{Ali} S.~S., {Bharadwaj} S., {Chengalur} J.~N., 2008, \mnras, 385, 2166

\bibitem[{{Ali} {et~al}\mbox{.}(2015){Ali}, {Parsons}, {Zheng}, {Pober}, {Liu},
  {Aguirre}, {Bradley}, {Bernardi}, {Carilli}, {Cheng}, {DeBoer}, {Dexter},
  {Grobbelaar}, {Horrell}, {Jacobs}, {Klima}, {MacMahon}, {Maree}, {Moore},
  {Razavi}, {Stefan}, {Walbrugh}, \& {Walker}}]{ali15}
{Ali} Z.~S. {et~al.}, 2015, \apj, 809, 61

\bibitem[{{Barkana}(2009)}]{barkana09}
{Barkana} R., 2009, \mnras, 397, 1454

\bibitem[{{Beardsley} {et~al}\mbox{.}(2013){Beardsley}, {Hazelton}, {Morales},
  {Arcus}, {Barnes}, {Bernardi}, {Bowman}, {Briggs}, {Bunton}, {Cappallo},
  {Corey}, {Deshpande}, {deSouza}, {Emrich}, {Gaensler}, {Goeke}, {Greenhill},
  {Herne}, {Hewitt}, {Johnston-Hollitt}, {Kaplan}, {Kasper}, {Kincaid},
  {Koenig}, {Kratzenberg}, {Lonsdale}, {Lynch}, {McWhirter}, {Mitchell},
  {Morgan}, {Oberoi}, {Ord}, {Pathikulangara}, {Prabu}, {Remillard}, {Rogers},
  {Roshi}, {Salah}, {Sault}, {Udaya}, {Srivani}, {Stevens}, {Subrahmanyan},
  {Tingay}, {Wayth}, {Waterson}, {Webster}, {Whitney}, {Williams}, {Williams},
  \& {Wyithe}}]{beardsley13}
{Beardsley} A.~P. {et~al.}, 2013, \mnras, 429, L5

\bibitem[{{Becker} {et~al}\mbox{.}(2015){Becker}, {Bolton}, {Madau}, {Pettini},
  {Ryan-Weber}, \& {Venemans}}]{becker15}
{Becker} G.~D., {Bolton} J.~S., {Madau} P., {Pettini} M., {Ryan-Weber} E.~V.,
  {Venemans} B.~P., 2015, \mnras, 447, 3402

\bibitem[{{Becker} {et~al}\mbox{.}(2001){Becker}, {Fan}, {White}, {Strauss},
  {Narayanan}, {Lupton}, {Gunn}, {Annis}, {Bahcall}, {Brinkmann}, {Connolly},
  {Csabai}, {Czarapata}, {Doi}, {Heckman}, {Hennessy}, {Ivezi{\'c}}, {Knapp},
  {Lamb}, {McKay}, {Munn}, {Nash}, {Nichol}, {Pier}, {Richards}, {Schneider},
  {Stoughton}, {Szalay}, {Thakar}, \& {York}}]{becker01}
{Becker} R.~H. {et~al.}, 2001, \aj, 122, 2850

\bibitem[{{Bernardi} {et~al}\mbox{.}(2009){Bernardi}, {de Bruyn}, {Brentjens},
  {Ciardi}, {Harker}, {Jeli{\'c}}, {Koopmans}, {Labropoulos}, {Offringa},
  {Pandey}, {Schaye}, {Thomas}, {Yatawatta}, \& {Zaroubi}}]{bernardi09}
{Bernardi} G. {et~al.}, 2009, \aap, 500, 965

\bibitem[{{Bharadwaj} \& {Ali}(2004)}]{bharadwaj04}
{Bharadwaj} S., {Ali} S.~S., 2004, \mnras, 352, 142

\bibitem[{{Bharadwaj} \& {Ali}(2005)}]{bharadwaj05}
{Bharadwaj} S., {Ali} S.~S., 2005, \mnras, 356, 1519

\bibitem[{{Bowman} {et~al}\mbox{.}(2013){Bowman}, {Cairns}, {Kaplan}, {Murphy},
  {Oberoi}, {Staveley-Smith}, {Arcus}, {Barnes}, {Bernardi}, {Briggs}, {Brown},
  {Bunton}, {Burgasser}, {Cappallo}, {Chatterjee}, {Corey}, {Coster},
  {Deshpande}, {deSouza}, {Emrich}, {Erickson}, {Goeke}, {Gaensler},
  {Greenhill}, {Harvey-Smith}, {Hazelton}, {Herne}, {Hewitt},
  {Johnston-Hollitt}, {Kasper}, {Kincaid}, {Koenig}, {Kratzenberg}, {Lonsdale},
  {Lynch}, {Matthews}, {McWhirter}, {Mitchell}, {Morales}, {Morgan}, {Ord},
  {Pathikulangara}, {Prabu}, {Remillard}, {Robishaw}, {Rogers}, {Roshi},
  {Salah}, {Sault}, {Shankar}, {Srivani}, {Stevens}, {Subrahmanyan}, {Tingay},
  {Wayth}, {Waterson}, {Webster}, {Whitney}, {Williams}, {Williams}, \&
  {Wyithe}}]{bowman13}
{Bowman} J.~D. {et~al.}, 2013, \pasa, 30, 31

\bibitem[{{Carron}, {Wolk} \& {Szapudi}(2015){Carron}, {Wolk}, \&
  {Szapudi}}]{caron14}
{Carron} J., {Wolk} M., {Szapudi} I., 2015, \mnras, 453, 450

\bibitem[{{Choudhury}, {Haehnelt} \& {Regan}(2009){Choudhury}, {Haehnelt}, \&
  {Regan}}]{choudhury09b}
{Choudhury} T.~R., {Haehnelt} M.~G., {Regan} J., 2009, \mnras, 394, 960

\bibitem[{{Davis} {et~al}\mbox{.}(1985){Davis}, {Efstathiou}, {Frenk}, \&
  {White}}]{davis85}
{Davis} M., {Efstathiou} G., {Frenk} C.~S., {White} S.~D.~M., 1985, \apj, 292,
  371

\bibitem[{{Di Matteo} {et~al}\mbox{.}(2002){Di Matteo}, {Perna}, {Abel}, \&
  {Rees}}]{dimatteo02}
{Di Matteo} T., {Perna} R., {Abel} T., {Rees} M.~J., 2002, \apj, 564, 576

\bibitem[{{Dillon} {et~al}\mbox{.}(2014){Dillon}, {Liu}, {Williams}, {Hewitt},
  {Tegmark}, {Morgan}, {Levine}, {Morales}, {Tingay}, {Bernardi}, {Bowman},
  {Briggs}, {Cappallo}, {Emrich}, {Mitchell}, {Oberoi}, {Prabu}, {Wayth}, \&
  {Webster}}]{dillon14}
{Dillon} J.~S. {et~al.}, 2014, \prd, 89, 023002

\bibitem[{{Dodelson}(2003)}]{dodelson_b03}
{Dodelson} S., 2003, {Modern cosmology}

\bibitem[{{Fan} {et~al}\mbox{.}(2003){Fan}, {Strauss}, {Schneider}, {Becker},
  {White}, {Haiman}, {Gregg}, {Pentericci}, {Grebel}, {Narayanan}, {Loh},
  {Richards}, {Gunn}, {Lupton}, {Knapp}, {Ivezi{\'c}}, {Brandt}, {Collinge},
  {Hao}, {Harbeck}, {Prada}, {Schaye}, {Strateva}, {Zakamska}, {Anderson},
  {Brinkmann}, {Bahcall}, {Lamb}, {Okamura}, {Szalay}, \& {York}}]{fan03}
{Fan} X. {et~al.}, 2003, \aj, 125, 1649

\bibitem[{{Feldman}, {Kaiser} \& {Peacock}(1994){Feldman}, {Kaiser}, \&
  {Peacock}}]{feldman94}
{Feldman} H.~A., {Kaiser} N., {Peacock} J.~A., 1994, \apj, 426, 23

\bibitem[{{Fialkov}, {Barkana} \& {Cohen}(2015){Fialkov}, {Barkana}, \&
  {Cohen}}]{fialkov15}
{Fialkov} A., {Barkana} R., {Cohen} A., 2015, Physical Review Letters, 114,
  101303

\bibitem[{{Furlanetto} {et~al}\mbox{.}(2009){Furlanetto}, {Lidz}, {Loeb},
  {McQuinn}, {Pritchard}, {Shapiro}, {Alvarez}, {Backer}, {Bowman}, {Burns},
  {Carilli}, {Cen}, {Cooray}, {Gnedin}, {Greenhill}, {Haiman}, {Hewitt},
  {Hirata}, {Lazio}, {Mesinger}, {Madau}, {Morales}, {Oh}, {Peterson},
  {Pihlstr{\"o}m}, {Tegmark}, {Trac}, {Zahn}, \& {Zaldarriaga}}]{furlanetto09}
{Furlanetto} S.~R. {et~al.}, 2009, in Astronomy, Vol. 2010, astro2010: The
  Astronomy and Astrophysics Decadal Survey, p.~82

\bibitem[{{Furlanetto}, {Zaldarriaga} \& {Hernquist}(2004){Furlanetto},
  {Zaldarriaga}, \& {Hernquist}}]{furlanetto04b}
{Furlanetto} S.~R., {Zaldarriaga} M., {Hernquist} L., 2004, \apj, 613, 1

\bibitem[{{Ghara}, {Choudhury} \& {Datta}(2015){Ghara}, {Choudhury}, \&
  {Datta}}]{ghara14}
{Ghara} R., {Choudhury} T.~R., {Datta} K.~K., 2015, \mnras, 447, 1806

\bibitem[{{Ghosh} {et~al}\mbox{.}(2012){Ghosh}, {Prasad}, {Bharadwaj}, {Ali},
  \& {Chengalur}}]{ghosh12}
{Ghosh} A., {Prasad} J., {Bharadwaj} S., {Ali} S.~S., {Chengalur} J.~N., 2012,
  \mnras, 426, 3295

\bibitem[{{Gleser}, {Nusser} \& {Benson}(2008){Gleser}, {Nusser}, \&
  {Benson}}]{gleser08}
{Gleser} L., {Nusser} A., {Benson} A.~J., 2008, \mnras, 391, 383

\bibitem[{{Goto} {et~al}\mbox{.}(2011){Goto}, {Utsumi}, {Hattori}, {Miyazaki},
  \& {Yamauchi}}]{goto11}
{Goto} T., {Utsumi} Y., {Hattori} T., {Miyazaki} S., {Yamauchi} C., 2011,
  \mnras, 415, L1

\bibitem[{{Iliev} {et~al}\mbox{.}(2014){Iliev}, {Mellema}, {Ahn}, {Shapiro},
  {Mao}, \& {Pen}}]{iliev13}
{Iliev} I.~T., {Mellema} G., {Ahn} K., {Shapiro} P.~R., {Mao} Y., {Pen} U.-L.,
  2014, \mnras, 439, 725

\bibitem[{{Jacobs} {et~al}\mbox{.}(2015){Jacobs}, {Pober}, {Parsons},
  {Aguirre}, {Ali}, {Bowman}, {Bradley}, {Carilli}, {DeBoer}, {Dexter},
  {Gugliucci}, {Klima}, {Liu}, {MacMahon}, {Manley}, {Moore}, {Stefan}, \&
  {Walbrugh}}]{jacobs14}
{Jacobs} D.~C. {et~al.}, 2015, \apj, 801, 51

\bibitem[{{Jeli{\'c}} {et~al}\mbox{.}(2008){Jeli{\'c}}, {Zaroubi},
  {Labropoulos}, {Thomas}, {Bernardi}, {Brentjens}, {de Bruyn}, {Ciardi},
  {Harker}, {Koopmans}, {Pandey}, {Schaye}, \& {Yatawatta}}]{jelic08}
{Jeli{\'c}} V. {et~al.}, 2008, \mnras, 389, 1319

\bibitem[{{Jensen} {et~al}\mbox{.}(2013){Jensen}, {Datta}, {Mellema},
  {Chapman}, {Abdalla}, {Iliev}, {Mao}, {Santos}, {Shapiro}, {Zaroubi},
  {Bernardi}, {Brentjens}, {de Bruyn}, {Ciardi}, {Harker}, {Jeli{\'c}},
  {Kazemi}, {Koopmans}, {Labropoulos}, {Martinez}, {Offringa}, {Pandey},
  {Schaye}, {Thomas}, {Veligatla}, {Vedantham}, \& {Yatawatta}}]{jensen13}
{Jensen} H. {et~al.}, 2013, \mnras, 435, 460

\bibitem[{{Komatsu} {et~al}\mbox{.}(2011){Komatsu}, {Smith}, {Dunkley},
  {Bennett}, {Gold}, {Hinshaw}, {Jarosik}, {Larson}, {Nolta}, {Page},
  {Spergel}, {Halpern}, {Hill}, {Kogut}, {Limon}, {Meyer}, {Odegard}, {Tucker},
  {Weiland}, {Wollack}, \& {Wright}}]{komatsu11}
{Komatsu} E. {et~al.}, 2011, \apjs, 192, 18

\bibitem[{{Koopmans} {et~al}\mbox{.}(2015){Koopmans}, {Pritchard}, {Mellema},
  {Aguirre}, {Ahn}, {Barkana}, {van Bemmel}, {Bernardi}, {Bonaldi}, {Briggs},
  {de Bruyn}, {Chang}, {Chapman}, {Chen}, {Ciardi}, {Dayal}, {Ferrara},
  {Fialkov}, {Fiore}, {Ichiki}, {Illiev}, {Inoue}, {Jelic}, {Jones}, {Lazio},
  {Maio}, {Majumdar}, {Mack}, {Mesinger}, {Morales}, {Parsons}, {Pen},
  {Santos}, {Schneider}, {Semelin}, {de Souza}, {Subrahmanyan}, {Takeuchi},
  {Vedantham}, {Wagg}, {Webster}, {Wyithe}, {Datta}, \& {Trott}}]{koopmans15}
{Koopmans} L. {et~al.}, 2015, Advancing Astrophysics with the Square Kilometre
  Array (AASKA14), 1

\bibitem[{{Lidz} {et~al}\mbox{.}(2008){Lidz}, {Zahn}, {McQuinn}, {Zaldarriaga},
  \& {Hernquist}}]{lidz08}
{Lidz} A., {Zahn} O., {McQuinn} M., {Zaldarriaga} M., {Hernquist} L., 2008,
  \apj, 680, 962

\bibitem[{{Majumdar}, {Bharadwaj} \& {Choudhury}(2013){Majumdar}, {Bharadwaj},
  \& {Choudhury}}]{majumdar13}
{Majumdar} S., {Bharadwaj} S., {Choudhury} T.~R., 2013, \mnras, 434, 1978

\bibitem[{{Majumdar} {et~al}\mbox{.}(2015){Majumdar}, {Jensen}, {Mellema},
  {Chapman}, {Abdalla}, {Lee}, {Iliev}, {Dixon}, {Datta}, {Ciardi},
  {Fernandez}, {Jeli{\'c}}, {Koopmans}, \& {Zaroubi}}]{majumdar15}
{Majumdar} S. {et~al.}, 2015, ArXiv e-prints: arXiv:1509.07518

\bibitem[{{Majumdar} {et~al}\mbox{.}(2014){Majumdar}, {Mellema}, {Datta},
  {Jensen}, {Choudhury}, {Bharadwaj}, \& {Friedrich}}]{majumdar14}
{Majumdar} S., {Mellema} G., {Datta} K.~K., {Jensen} H., {Choudhury} T.~R.,
  {Bharadwaj} S., {Friedrich} M.~M., 2014, \mnras, 443, 2843

\bibitem[{{McQuinn} {et~al}\mbox{.}(2007){McQuinn}, {Lidz}, {Zahn}, {Dutta},
  {Hernquist}, \& {Zaldarriaga}}]{mcquinn07}
{McQuinn} M., {Lidz} A., {Zahn} O., {Dutta} S., {Hernquist} L., {Zaldarriaga}
  M., 2007, \mnras, 377, 1043

\bibitem[{{McQuinn} {et~al}\mbox{.}(2006){McQuinn}, {Zahn}, {Zaldarriaga},
  {Hernquist}, \& {Furlanetto}}]{mcquinn06}
{McQuinn} M., {Zahn} O., {Zaldarriaga} M., {Hernquist} L., {Furlanetto} S.~R.,
  2006, \apj, 653, 815

\bibitem[{{Mellema} {et~al}\mbox{.}(2013){Mellema}, {Koopmans}, {Abdalla},
  {Bernardi}, {Ciardi}, {Daiboo}, {de Bruyn}, {Datta}, {Falcke}, {Ferrara},
  {Iliev}, {Iocco}, {Jeli{\'c}}, {Jensen}, {Joseph}, {Labroupoulos}, {Meiksin},
  {Mesinger}, {Offringa}, {Pandey}, {Pritchard}, {Santos}, {Schwarz},
  {Semelin}, {Vedantham}, {Yatawatta}, \& {Zaroubi}}]{mellema13}
{Mellema} G. {et~al.}, 2013, Experimental Astronomy, 36, 235

\bibitem[{{Mesinger}, {Furlanetto} \& {Cen}(2011){Mesinger}, {Furlanetto}, \&
  {Cen}}]{mesinger11}
{Mesinger} A., {Furlanetto} S., {Cen} R., 2011, \mnras, 411, 955

\bibitem[{{Mitra}, {Choudhury} \& {Ferrara}(2011){Mitra}, {Choudhury}, \&
  {Ferrara}}]{mitra11}
{Mitra} S., {Choudhury} T.~R., {Ferrara} A., 2011, \mnras, 413, 1569

\bibitem[{{Mitra}, {Choudhury} \& {Ferrara}(2015){Mitra}, {Choudhury}, \&
  {Ferrara}}]{mitra15}
{Mitra} S., {Choudhury} T.~R., {Ferrara} A., 2015, ArXiv e-prints,
  arXiv:1505.05507

\bibitem[{{Mitra}, {Ferrara} \& {Choudhury}(2013){Mitra}, {Ferrara}, \&
  {Choudhury}}]{mitra13}
{Mitra} S., {Ferrara} A., {Choudhury} T.~R., 2013, \mnras, 428, L1

\bibitem[{{Mohammed} \& {Seljak}(2014)}]{mohammed14}
{Mohammed} I., {Seljak} U., 2014, \mnras, 445, 3382

\bibitem[{{Mondal} {et~al}\mbox{.}(2015){Mondal}, {Bharadwaj}, {Majumdar},
  {Bera}, \& {Acharyya}}]{mondal15}
{Mondal} R., {Bharadwaj} S., {Majumdar} S., {Bera} A., {Acharyya} A., 2015,
  \mnras, 449, L41

\bibitem[{{Moore} {et~al}\mbox{.}(2015){Moore}, {Aguirre}, {Parsons}, {Ali},
  {Bradley}, {Carilli}, {DeBoer}, {Dexter}, {Gugliucci}, {Jacobs}, {Klima},
  {Liu}, {MacMahon}, {Manley}, {Pober}, {Stefan}, \& {Walbrugh}}]{moore15}
{Moore} D. {et~al.}, 2015, ArXiv e-prints, arXiv:1502.05072

\bibitem[{{Moore} {et~al}\mbox{.}(2013){Moore}, {Aguirre}, {Parsons}, {Jacobs},
  \& {Pober}}]{moore13}
{Moore} D.~F., {Aguirre} J.~E., {Parsons} A.~R., {Jacobs} D.~C., {Pober} J.~C.,
  2013, \apj, 769, 154

\bibitem[{{Morales}(2005)}]{morales05}
{Morales} M.~F., 2005, \apj, 619, 678

\bibitem[{{Neyrinck}(2011)}]{neyrinck11}
{Neyrinck} M.~C., 2011, \apj, 736, 8

\bibitem[{{Paciga} {et~al}\mbox{.}(2013){Paciga}, {Albert}, {Bandura}, {Chang},
  {Gupta}, {Hirata}, {Odegova}, {Pen}, {Peterson}, {Roy}, {Shaw}, {Sigurdson},
  \& {Voytek}}]{paciga13}
{Paciga} G. {et~al.}, 2013, \mnras, 433, 639

\bibitem[{{Parsons} {et~al}\mbox{.}(2014){Parsons}, {Liu}, {Aguirre}, {Ali},
  {Bradley}, {Carilli}, {DeBoer}, {Dexter}, {Gugliucci}, {Jacobs}, {Klima},
  {MacMahon}, {Manley}, {Moore}, {Pober}, {Stefan}, \& {Walbrugh}}]{parsons14}
{Parsons} A.~R. {et~al.}, 2014, \apj, 788, 106

\bibitem[{{Patil} {et~al}\mbox{.}(2014){Patil}, {Zaroubi}, {Chapman},
  {Jeli{\'c}}, {Harker}, {Abdalla}, {Asad}, {Bernardi}, {Brentjens}, {de
  Bruyn}, {Bus}, {Ciardi}, {Daiboo}, {Fernandez}, {Ghosh}, {Jensen}, {Kazemi},
  {Koopmans}, {Labropoulos}, {Mevius}, {Martinez}, {Mellema}, {Offringa},
  {Pandey}, {Schaye}, {Thomas}, {Vedantham}, {Veligatla}, {Wijnholds}, \&
  {Yatawatta}}]{patil14}
{Patil} A.~H. {et~al.}, 2014, \mnras, 443, 1113

\bibitem[{{Peebles}(1980)}]{peebles1980}
{Peebles} P.~J.~E., 1980, {The large-scale structure of the universe}

\bibitem[{{Planck Collaboration}(2015)}]{planck15}
{Planck Collaboration}, 2015, ArXiv e-prints, arXiv:1502.01589

\bibitem[{{Planck Collaboration} {et~al}\mbox{.}(2014){Planck Collaboration},
  {Ade}, {Aghanim}, {Armitage-Caplan}, {Arnaud}, {Ashdown}, {Atrio-Barandela},
  {Aumont}, {Baccigalupi}, {Banday}, \& et~al.}]{planck14}
{Planck Collaboration} {et~al.}, 2014, \aap, 571, A16

\bibitem[{{Pober} {et~al}\mbox{.}(2014){Pober}, {Liu}, {Dillon}, {Aguirre},
  {Bowman}, {Bradley}, {Carilli}, {DeBoer}, {Hewitt}, {Jacobs}, {McQuinn},
  {Morales}, {Parsons}, {Tegmark}, \& {Werthimer}}]{pober14}
{Pober} J.~C. {et~al.}, 2014, \apj, 782, 66

\bibitem[{{Pober} {et~al}\mbox{.}(2013){Pober}, {Parsons}, {Aguirre}, {Ali},
  {Bradley}, {Carilli}, {DeBoer}, {Dexter}, {Gugliucci}, {Jacobs}, {Klima},
  {MacMahon}, {Manley}, {Moore}, {Stefan}, \& {Walbrugh}}]{pober13}
{Pober} J.~C. {et~al.}, 2013, \apjl, 768, L36

\bibitem[{{Robertson} {et~al}\mbox{.}(2015){Robertson}, {Ellis}, {Furlanetto},
  \& {Dunlop}}]{robertson15}
{Robertson} B.~E., {Ellis} R.~S., {Furlanetto} S.~R., {Dunlop} J.~S., 2015,
  \apjl, 802, L19

\bibitem[{{Robertson} {et~al}\mbox{.}(2013){Robertson}, {Furlanetto},
  {Schneider}, {Charlot}, {Ellis}, {Stark}, {McLure}, {Dunlop}, {Koekemoer},
  {Schenker}, {Ouchi}, {Ono}, {Curtis-Lake}, {Rogers}, {Bowler}, \&
  {Cirasuolo}}]{robertson13}
{Robertson} B.~E. {et~al.}, 2013, \apj, 768, 71

\bibitem[{{Shapiro} {et~al}\mbox{.}(2013){Shapiro}, {Mao}, {Iliev}, {Mellema},
  {Datta}, {Ahn}, \& {Koda}}]{shapiro13}
{Shapiro} P.~R., {Mao} Y., {Iliev} I.~T., {Mellema} G., {Datta} K.~K., {Ahn}
  K., {Koda} J., 2013, Physical Review Letters, 110, 151301

\bibitem[{{Tingay} {et~al}\mbox{.}(2013){Tingay}, {Goeke}, {Bowman}, {Emrich},
  {Ord}, {Mitchell}, {Morales}, {Booler}, {Crosse}, {Wayth}, {Lonsdale},
  {Tremblay}, {Pallot}, {Colegate}, {Wicenec}, {Kudryavtseva}, {Arcus},
  {Barnes}, {Bernardi}, {Briggs}, {Burns}, {Bunton}, {Cappallo}, {Corey},
  {Deshpande}, {Desouza}, {Gaensler}, {Greenhill}, {Hall}, {Hazelton}, {Herne},
  {Hewitt}, {Johnston-Hollitt}, {Kaplan}, {Kasper}, {Kincaid}, {Koenig},
  {Kratzenberg}, {Lynch}, {Mckinley}, {Mcwhirter}, {Morgan}, {Oberoi},
  {Pathikulangara}, {Prabu}, {Remillard}, {Rogers}, {Roshi}, {Salah}, {Sault},
  {Udaya-Shankar}, {Schlagenhaufer}, {Srivani}, {Stevens}, {Subrahmanyan},
  {Waterson}, {Webster}, {Whitney}, {Williams}, {Williams}, \&
  {Wyithe}}]{tingay13}
{Tingay} S.~J. {et~al.}, 2013, \pasa, 30, 7

\bibitem[{{van Haarlem} {et~al}\mbox{.}(2013){van Haarlem}, {Wise}, {Gunst},
  {Heald}, {McKean}, {Hessels}, {de Bruyn}, {Nijboer}, {Swinbank}, {Fallows},
  {Brentjens}, {Nelles}, {Beck}, {Falcke}, {Fender}, {H{\"o}randel},
  {Koopmans}, {Mann}, {Miley}, {R{\"o}ttgering}, {Stappers}, {Wijers},
  {Zaroubi}, {van den Akker}, {Alexov}, {Anderson}, {Anderson}, {van Ardenne},
  {Arts}, {Asgekar}, {Avruch}, {Batejat}, {B{\"a}hren}, {Bell}, {Bell}, {van
  Bemmel}, {Bennema}, {Bentum}, {Bernardi}, {Best}, {B{\^i}rzan}, {Bonafede},
  {Boonstra}, {Braun}, {Bregman}, {Breitling}, {van de Brink}, {Broderick},
  {Broekema}, {Brouw}, {Br{\"u}ggen}, {Butcher}, {van Cappellen}, {Ciardi},
  {Coenen}, {Conway}, {Coolen}, {Corstanje}, {Damstra}, {Davies}, {Deller},
  {Dettmar}, {van Diepen}, {Dijkstra}, {Donker}, {Doorduin}, {Dromer}, {Drost},
  {van Duin}, {Eisl{\"o}ffel}, {van Enst}, {Ferrari}, {Frieswijk}, {Gankema},
  {Garrett}, {de Gasperin}, {Gerbers}, {de Geus}, {Grie{\ss}meier}, {Grit},
  {Gruppen}, {Hamaker}, {Hassall}, {Hoeft}, {Holties}, {Horneffer}, {van der
  Horst}, {van Houwelingen}, {Huijgen}, {Iacobelli}, {Intema}, {Jackson},
  {Jelic}, {de Jong}, {Juette}, {Kant}, {Karastergiou}, {Koers}, {Kollen},
  {Kondratiev}, {Kooistra}, {Koopman}, {Koster}, {Kuniyoshi}, {Kramer},
  {Kuper}, {Lambropoulos}, {Law}, {van Leeuwen}, {Lemaitre}, {Loose}, {Maat},
  {Macario}, {Markoff}, {Masters}, {McFadden}, {McKay-Bukowski}, {Meijering},
  {Meulman}, {Mevius}, {Middelberg}, {Millenaar}, {Miller-Jones}, {Mohan},
  {Mol}, {Morawietz}, {Morganti}, {Mulcahy}, {Mulder}, {Munk}, {Nieuwenhuis},
  {van Nieuwpoort}, {Noordam}, {Norden}, {Noutsos}, {Offringa}, {Olofsson},
  {Omar}, {Orr{\'u}}, {Overeem}, {Paas}, {Pandey-Pommier}, {Pandey}, {Pizzo},
  {Polatidis}, {Rafferty}, {Rawlings}, {Reich}, {de Reijer}, {Reitsma},
  {Renting}, {Riemers}, {Rol}, {Romein}, {Roosjen}, {Ruiter}, {Scaife}, {van
  der Schaaf}, {Scheers}, {Schellart}, {Schoenmakers}, {Schoonderbeek},
  {Serylak}, {Shulevski}, {Sluman}, {Smirnov}, {Sobey}, {Spreeuw}, {Steinmetz},
  {Sterks}, {Stiepel}, {Stuurwold}, {Tagger}, {Tang}, {Tasse}, {Thomas},
  {Thoudam}, {Toribio}, {van der Tol}, {Usov}, {van Veelen}, {van der Veen},
  {ter Veen}, {Verbiest}, {Vermeulen}, {Vermaas}, {Vocks}, {Vogt}, {de Vos},
  {van der Wal}, {van Weeren}, {Weggemans}, {Weltevrede}, {White}, {Wijnholds},
  {Wilhelmsson}, {Wucknitz}, {Yatawatta}, {Zarka}, {Zensus}, \& {van
  Zwieten}}]{haarlem13}
{van Haarlem} M.~P. {et~al.}, 2013, \aap, 556, A2

\bibitem[{{Watkinson} \& {Pritchard}(2014)}]{watkinson14}
{Watkinson} C.~A., {Pritchard} J.~R., 2014, \mnras, 443, 3090

\bibitem[{{White} {et~al}\mbox{.}(2003){White}, {Becker}, {Fan}, \&
  {Strauss}}]{white03}
{White} R.~L., {Becker} R.~H., {Fan} X., {Strauss} M.~A., 2003, \aj, 126, 1

\bibitem[{{Yatawatta} {et~al}\mbox{.}(2013){Yatawatta}, {de Bruyn},
  {Brentjens}, {Labropoulos}, {Pandey}, {Kazemi}, {Zaroubi}, {Koopmans},
  {Offringa}, {Jeli{\'c}}, {Martinez Rubi}, {Veligatla}, {Wijnholds}, {Brouw},
  {Bernardi}, {Ciardi}, {Daiboo}, {Harker}, {Mellema}, {Schaye}, {Thomas},
  {Vedantham}, {Chapman}, {Abdalla}, {Alexov}, {Anderson}, {Avruch}, {Batejat},
  {Bell}, {Bell}, {Bentum}, {Best}, {Bonafede}, {Bregman}, {Breitling}, {van de
  Brink}, {Broderick}, {Br{\"u}ggen}, {Conway}, {de Gasperin}, {de Geus},
  {Duscha}, {Falcke}, {Fallows}, {Ferrari}, {Frieswijk}, {Garrett},
  {Griessmeier}, {Gunst}, {Hassall}, {Hessels}, {Hoeft}, {Iacobelli}, {Juette},
  {Karastergiou}, {Kondratiev}, {Kramer}, {Kuniyoshi}, {Kuper}, {van Leeuwen},
  {Maat}, {Mann}, {McKean}, {Mevius}, {Mol}, {Munk}, {Nijboer}, {Noordam},
  {Norden}, {Orru}, {Paas}, {Pandey-Pommier}, {Pizzo}, {Polatidis}, {Reich},
  {R{\"o}ttgering}, {Sluman}, {Smirnov}, {Stappers}, {Steinmetz}, {Tagger},
  {Tang}, {Tasse}, {ter Veen}, {Vermeulen}, {van Weeren}, {Wise}, {Wucknitz},
  \& {Zarka}}]{yatawatta13}
{Yatawatta} S. {et~al.}, 2013, \aap, 550, A136

\end{thebibliography}
\bsp

\label{lastpage}

\end{document}